\documentclass[aps,twocolumn,
superscriptaddress
]{revtex4}
\pdfoutput=0

\usepackage[latin1]{inputenc}
\usepackage{amsmath}
\usepackage{amsfonts}
\usepackage{amssymb}
\usepackage{graphicx}
 \usepackage{color}

\newcommand{\defeq}{:=}
\newcommand{\eqdef}{=:}
 
\newcommand{\bra}[1]{\langle #1|}
\newcommand{\ket}[1]{\left|#1\right\rangle}

\newcommand{\abs}[1]{{\left\vert#1\right\vert}}

\newcommand{\kb}[2]{\ensuremath{\vert #1 \rangle \langle #2 \vert}}

\newcommand{\id}[0]{\ensuremath{\mathbf{1}}}

\begin{document}
\title{Superabsorption of light via quantum engineering}
\author{K. D. B. Higgins}
\email{kieran.higgins@materials.ox.ac.uk}
\affiliation{Department of Materials, Oxford University, Oxford OX1 3PH, United Kingdom}
\author{S. C. Benjamin}
\affiliation{Department of Materials, Oxford University, Oxford OX1 3PH, United Kingdom}
\affiliation{Centre for Quantum Technologies, National University of Singapore, 3 Science Drive 2, Singapore 117543}
\author{T. M. Stace}
\affiliation{Centre for Engineered Quantum Systems, School of Mathematics and Physics, The University of Queensland,
St Lucia, Queensland 4072, Australia}
\author{G. J. Milburn}
\affiliation{Centre for Engineered Quantum Systems, School of Mathematics and Physics, The University of Queensland,
St Lucia, Queensland 4072, Australia}
\author{B. W. Lovett}
\affiliation{SUPA, School of Physics and Astronomy, University of St Andrews, KY16 9SS, United Kingdom}
\affiliation{Department of Materials, Oxford University, Oxford OX1 3PH, United Kingdom}
\author{E. M. Gauger}
\email{erik.gauger@nus.edu.sg}
\affiliation{Centre for Quantum Technologies, National University of Singapore, 3 Science Drive 2, Singapore 117543}
\affiliation{Department of Materials, Oxford University, Oxford OX1 3PH, United Kingdom}
%
\begin{abstract}

Almost 60 years ago Dicke introduced the term superradiance to describe a signature quantum effect: $N$ atoms can collectively emit light at a rate proportional to  $N^2$. Structures that superradiate must also have enhanced absorption, but the former always dominates in natural systems. Here we show that this restriction can be overcome by combining several well-established quantum control techniques. Our analytical and numerical calculations show that superabsorption can then be achieved and sustained in certain simple nanostructures, by trapping the system in a highly excited state through transition rate engineering. This opens the prospect of a new class of quantum nanotechnology with potential applications including photon detection and light-based power transmission. An array of quantum dots or a molecular ring structure could provide a suitable platform for an experimental demonstration.
\end{abstract}

\maketitle

Superradiance can occur when $N$ individual atoms interact with the surrounding electromagnetic field \cite{dicke1954}. 
Here we use the term `atom' broadly to refer to entities with a discrete dipole-allowed transition, including semiconductor quantum dots \cite{scheibner2007}, crystal defects, and molecules \cite{wang1995a}.
Following an initial excitation of all atoms, dipole-allowed decay down a series of symmetrical `Dicke ladder' states leads to an enhanced light-matter coupling that, when the system reaches the state half way down the ladder, depends on the square of the atomic transition dipole. Thus when $N$ dipoles add coherently, light can be emitted at an enhanced rate proportional to $N^2$~\cite{dicke1954, gross1982, brandes2005}. 
Even for moderate $N$ this represents a significant increase over the prediction of classical physics, and the effect has found applications ranging from probing exciton delocalisation in biological systems \cite{monshouwer1997}, to developing a new class of laser \cite{bohnet2012}, and may even lead to observable effects in astrophysics \cite{putten1999}.

Time-reversal symmetry of quantum mechanics implies that systems with enhanced emission rates will also have enhanced absorption rates. Naturally emission dominates if an excited state of the collective emits into a vacuum, since there are no photons to absorb. Even in an intense light field where absorption and emission are closely balanced, a given transition remains more likely to emit than to absorb. Thus it might seem that the inverse of superradiance is intrinsically ephemeral.

\begin{figure}[b]
\includegraphics[width=0.87\columnwidth]{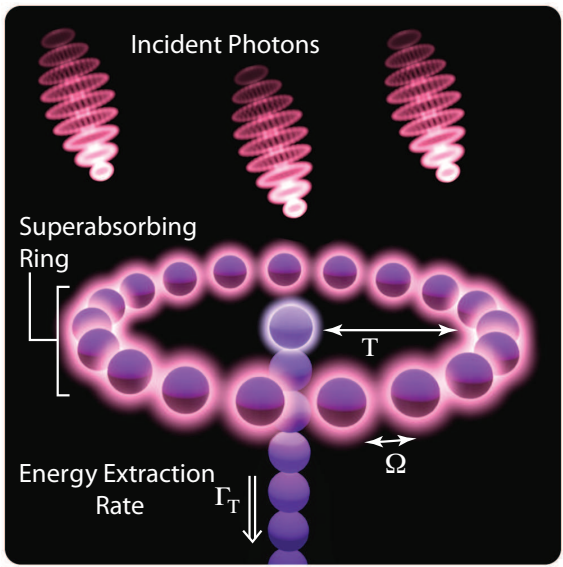}
\caption{\label{fig:schematic} One potential realisation of superabsorption: Photons absorbed by the ring give rise to delocalised excitons; ideally the ring maintains a specific exciton population to achieve enhanced absorption. Combined with a suitable charge sensor (e.g. a quantum point contact) this enables photon sensing. We also model an application for photon harvesting, where newly created excitons are transferred from the ring to a central core absorber, followed by an irreversible process (e.g. one-way transfer down a strongly coupled chain) to a centre converting the exciton into stored energy.}
\end{figure}

However, here we show that certain interactions between the atoms allow us to control a quantum system such that a sustained superabsorbing state can exist. For atoms in close proximity and with a suitable geometrical arrangement, ever present atomic dipolar interactions are sufficient for our purposes. An appropriate realisation involves a ring structure that is strikingly reminiscent of the photosynthetic light harvesting complex LH1 \cite{blankenship02, dong2012}  (see Fig.~\ref{fig:schematic}). Although the potential for enhanced absorption inherently exists in all superradiating systems, natural systems are not designed to ulitise it. Rather, these will always perform an (often strongly) biased random walk down the ladder of accessible states, being attracted by the bottom most rung. Strongly enhanced absorption near the middle of the Dicke ladder is thus an improbable process and can only last for a vanishingly short time. 

By contrast, in this Communication, we will show how to harness {\it environmental quantum control} techniques to break the dominance of emission over absorption and extend the time during which a collective system maintains the capability for quantum enhanced absorption.  By interfacing the well-established physical phenomena of superradiance, light filtering, photonic band gaps, and quantum feedback control, we show that sustained superlinear scaling of the light absorption rate with the number of atoms is possible. Since this represents the reciprocal process to superradiance, we shall refer to it as `superabsorption'. Note that this effect is quite distinct from other recent studies investigating  collective light-matter interactions in the context of  `cloaking'  \cite{chen2013} and time-reversed lasing \cite{wan2011}. 

In the following we will present the Dicke model of superradiance before describing the requirements for unlocking engineered superabsorption. Our discussion will explore its potential for practical technologies through the examples of photon sensing and light-based energy transmission.
 
\section{Results}

\subsection{Superradiance}

The Hamiltonian of an ensemble of $N$ identical atoms is ($\hbar=1$):
\begin{equation}\label{Hamiltonian}
\hat{H}_{S} = \frac{\omega_{A}}{2} \sum_{i=1}^{N} \left( \id^{i} - \hat{\sigma}^{i}_{z} \right) =  \omega_{A} \sum_{i=1}^{N} \hat{\sigma}^{i}_{+}\hat{\sigma}^{i}_{-} \,,
\end{equation}
where $\omega_{A}$ is the bare atomic transition frequency; $\hat{\sigma}^{i}_{z} = \ket{g}_{i}\bra{g} - \ket{e}_{i}\bra{e}$, $\hat{\sigma}^{i}_{+} = \ket{e}_{i}\bra{g}$, and $\hat{\sigma}^{i}_{-} = (\hat{\sigma}^{i}_{+})^{\dagger}$  are the usual Pauli operators defined with respect to the $i$th atom's ground, $\ket{g}_{i}$, and optically excited state, $\ket{e}_{i}$. When the wavelength $\lambda$ of light is much larger than all interatomic distances $r_{ij}$, $(\lambda \gg r_{ij})$, the atoms become indistinguishable and light interacts with the system collectively. The dynamics are then best described by collective operators:
\begin{figure*}
\includegraphics[scale=0.75]{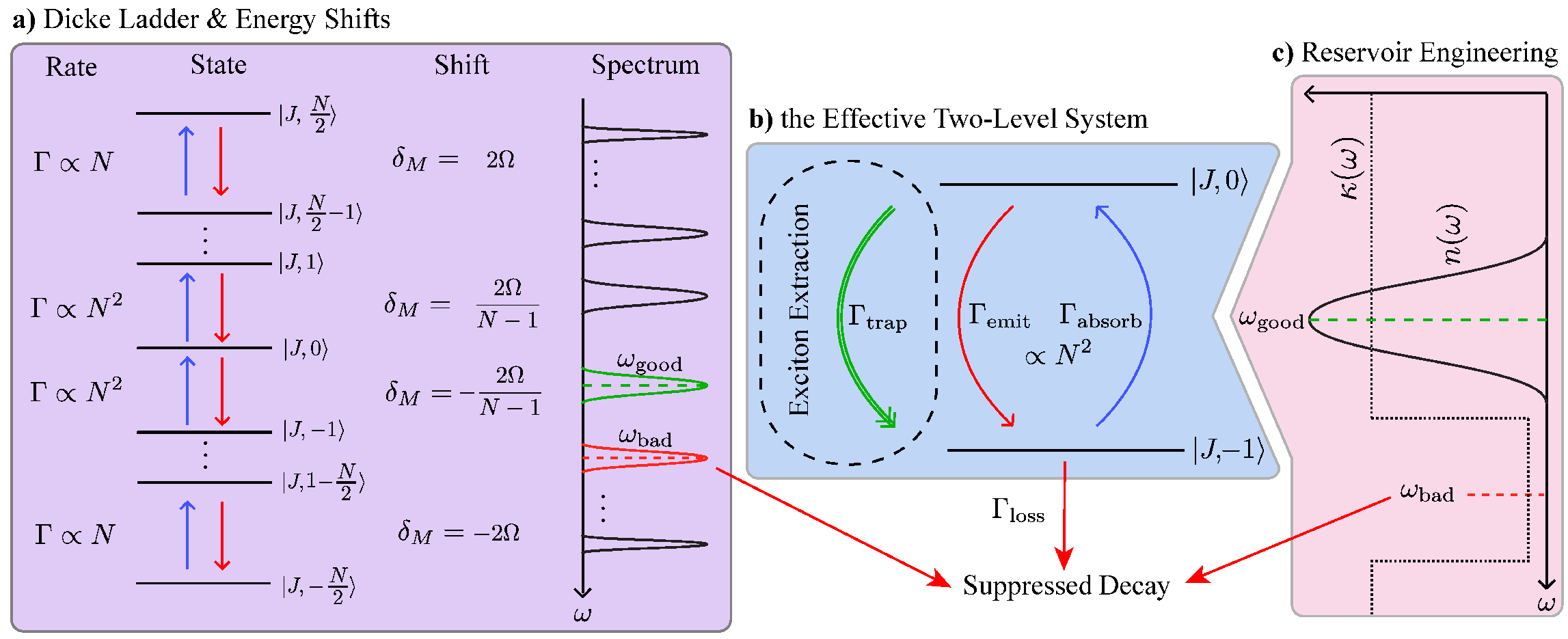}
\caption{\label{fig:1} \textbf{Left:} The ladder of Dicke states of an $N$ atom system, with emission (red) and absorption (blue) processes. In the presence of interactions $\Omega \not= 0$, the frequency shift of each transition is given by $\omega_{A} + \delta_M$. \textbf{Middle:} The Effective Two-Level System (E2LS) picture with optional trapping process for energy extraction in the dashed box. \textbf{Right:} A scheme for using the environment to confine the ladder of states into an effective two level system either by tailoring the spectral density $\kappa(\omega)$ or the mode occupation $n(\omega)$.}
\end{figure*}
\begin{equation}\label{Eqn_opers}
\hat{J}_{\pm} = \sum_{i=1}^{N} \hat{\sigma}_{\pm}^{i},~~ ~~ \hat{J}_{z} = \sum_{i=1}^{N} \hat{\sigma}_{z}^{i},
\end{equation}
which generate transitions between the eigenstates of the Hamiltonian~(\ref{Hamiltonian}) and obey $SU(2)$ commutation relations. We can succinctly express the light matter interaction Hamiltonian as
\begin{equation}\label{interaction_L}
\hat{H}_{L}= -\hat{E} \,d \left( \hat{J}_{+} +\hat{J}_{-} \right) \, ,
\end{equation} 
where $\hat{E}$ is the light field operator and $d$ is the atomic dipole matrix element. The Hamiltonian~(\ref{interaction_L}) causes the system to move along a ladder of states called the `Dicke' or `bright' states which are characterised by the eigenvalues $J$ and $M$ of $\hat{J}^2$ and $\hat{J}_{z}$, respectively. In the absence of interactions between the atoms, $\hat{J}^2$ commutes with $\hat H_{S}+\hat H_{L}$ and thus its eigenvalue $\frac{N}{2} \left(\frac{N}{2}+1 \right)$ is a conserved quantity. The Dicke states form a ladder from $\ket{J,-\frac{N}{2}}$ to  $\ket{J,\frac{N}{2}}$ shown in Fig.~\ref{fig:1}a; the $N+1$ rungs correspond to the fully symmetric superpositions of $N/2+M$ excited atoms for each value of $M$. The collective excitation operators %
\begin{equation}\label{eq:SRmatrix}
\hat{J}_{\pm}\ket{J,M}=\sqrt{(J \pm M+1)(J \mp M)}\ket{J,M+1},
\end{equation}
explore this ladder of states,  and the transition rates between adjacent Dicke ladder states are then readily calculated:
\begin{equation}\label{SRrates}
\Gamma_{M\rightarrow M\pm1}=\gamma \left( \frac{N}{2}\pm M+1 \right) \left( \frac{N}{2} \mp M \right),
\end{equation}
where $\gamma= 8\pi^2 d^{2} / \left( 3 \epsilon_{0} \hbar \lambda^{3} \right)$ is the free atom decay rate. 

If the system is initialized in the fully excited state $\ket{J,\frac{N}{2}}$ with no environmental photons, then the system cascades down the ladder, as shown by the red arrows in Fig.~\ref{fig:1}a. Upon reaching the midway point $(M=0)$ its emission rate {exceeds the rate $\gamma N$ expected of $N$ uncorrelated atoms for $N > 2$. For a larger number of atoms the peak transition rate of Eq.~(\ref{SRrates}) follows a quadratic dependence on $N$ and is well approximated by
\begin{equation}\label{SRratesm0}
\Gamma_{M\rightarrow M-1}\approx \gamma \left( \frac{N}{2} \right)^2 \, . 
\end{equation}
This is the essence of superradiance: constructive interference between the different possible decay paths greatly enhances the emission rate, producing a high intensity pulse. The enhancement is the result of simple combinatorics: near the middle of the ladder, $\ket{J,0}$, there are a large number of possible configurations of excited atoms that contribute to each respective Dicke state. 

Superradiance is not an intrinsically transient effect: steady state operation can occur through repumping \cite{haake1993}, or in cavities \cite{meiser2010, auffeves2011}, and recently a superradiant laser with potential for extraordinary stability and narrow linewidth has been demonstrated \cite{bohnet2012}.

\subsection{Superabsorption}

The crucial ingredient for achieving superabsorption is to engineer the transition rates in a way that primarily confines the dynamics  to an {\it effective two-level system} (E2LS) around the $M = 0$ transition (see Fig.~\ref{fig:1}b), which exhibits the required quadratic absorption rate as depicted in the inset of Fig.~\ref{fig:new}. 

In order to ensure that most transitions take place within the E2LS we must either suppress the total loss rate from the E2LS or enhance the probability of transitions within it. This becomes possible if the frequency of the E2LS transition is distinct from that  of other transitions, and in particular the one immediately below the targeted transition within the E2LS. This will never be the case for a non-interacting set of atoms, which must have a degenerate set of ladder transition energies, but it can occur once suitable interactions are included. Dicke physics requires that the atoms remain identical but interactions are still permissible in certain symmetric geometries such as rings \cite{coffey1978,gross1982}, and these structures will continue to exhibit superradiance, and are therefore also capable of superabsorption.

To show this, we consider the candidate superabsorber depicted in Fig.~\ref{fig:schematic}. We assume the interactions act between adjacent atoms only and are due to F\"orster type coupling. This leads to a Dicke ladder of non-degenerate transitions whose dynamics are found from a collective quantum optical master equation:
\begin{align}\label{MEgen}
&\dot{\rho}  =	 - i [\hat{H}_S+\hat{H}_{I}, \rho]  \\
&			 - \gamma \sum_{\beta \in \omega} \kappa(\omega_{\beta}) \left( (n(\omega_{\beta})+1) D[\hat{L}_{\beta}]\rho + n(\omega_{\beta}) D[\hat{L}^{\dagger}_{\beta}]\rho \right) \, . \nonumber
\end{align}
$\kappa(\omega)= \sum_{k}\abs{g_{k}}^2\delta(\omega-\omega_{k})= \chi({\omega})\abs{g(\omega)}^{2}$ is the spectral density at frequency $\omega$; $n(\omega_{\beta})$ is the occupation number of the $\omega_{\beta}$ mode, and $D[\hat{L}_{\beta}]\rho$ is the Lindbladian dissipator $\hat{L}_{\beta}\rho\hat{L}^{\dagger}_{\beta}-\frac{1}{2}\{\hat{L}^{\dagger}_{\beta}\hat{L}_{\beta},\rho\}$. $L_\beta^\dagger$ moves the system up a Dicke ladder transition with frequency $\omega_\beta$. 

Eq.~(\ref{MEgen}) also features unitary dynamics due to the field interaction that comprises two components: the Lamb shift, accounted for by renormalising $\omega_{A}$ in the system Hamiltonian $\hat H_S$,  
and the field induced dipole-dipole interaction
\begin{equation}\label{Hopping}
\hat{H}_{I}= \Omega_{i,j} \sum_{i \neq j}^{N} \left( \hat{\sigma}_{+}^{i} \hat{\sigma}_{-}^{j}+ \hat{\sigma}_{-}^{i} \hat{\sigma}_{+}^{j} \right) \, ,
\end{equation}
which describes energy conserving `hopping' of excitons between sites mediated by virtual photon exchange. The hopping interaction strength $\Omega_{i,j}$  is given by \cite{gross1982}
\begin{equation}\label{eq:omega}
\Omega(i,j)= \frac{d^2}{4\pi \epsilon_{0} r^{3}_{ij}} \left[1-\frac{3(\hat{\epsilon}_{a} \cdot \vec{r}_{ij})^{2}}{r^{2}_{ij}}\right]\approx \frac{d^2}{4\pi \epsilon_{0} r_{ij}^{3}} 
\end{equation}
with $\hat\epsilon_{a}$ being a unit vector parallel to the direction of the dipoles. For a circular geometry with dipoles perpendicular to $\vec{r}_{ij}$ and retaining only nearest neighbour interactions [a good approximation for larger rings since $\Omega(i,j)\propto r_{ij}^{-3}$], $\Omega \defeq \Omega(i,i+1)$ is a constant. However, note that the restriction to nearest neighbour coupling is not a requirement, please see the Supplementary Information for a full discussion. Owing to the high degree of symmetry of the ring geometry, to first order $\hat{H}_I$ does not mix the $\ket{JM}$ eigenstates, only shifting their energies \cite{gross1982} according to
\begin{equation}\label{eq:energy_shift}
\delta E_{M}=\bra{J,M} \hat{H}_{I}\ket{J,M}=\Omega \frac{J^{2}-M^{2}}{J-\frac{1}{2}} \, .
\end{equation}
The shift of the transition frequencies is given by the difference of two adjacent levels $ E_{M}-E_{M-1}$:
\begin{equation}\label{shift}
\omega_{M \rightarrow M-1}=\omega_{A}-4\Omega\frac{M-\frac{1}{2}}{N-1} \, .
\end{equation}
These altered frequencies break the degeneracy in the Dicke ladder where each transition now has a unique frequency. For example the transition frequency from the ground state to the first Dicke state is $\omega_{-N/2+1 \rightarrow -N/2}=\omega_{A}-2\Omega$. Crucially, the Dicke states still still represent a very good approximation of the eigenbasis of the system, yet each transition in the ladder now samples both $\kappa(\omega)$ and $n(\omega)$ at its own unique frequency. 

\subsection{Transition Rate Engineering}

\begin{figure}
\includegraphics[width=\linewidth]{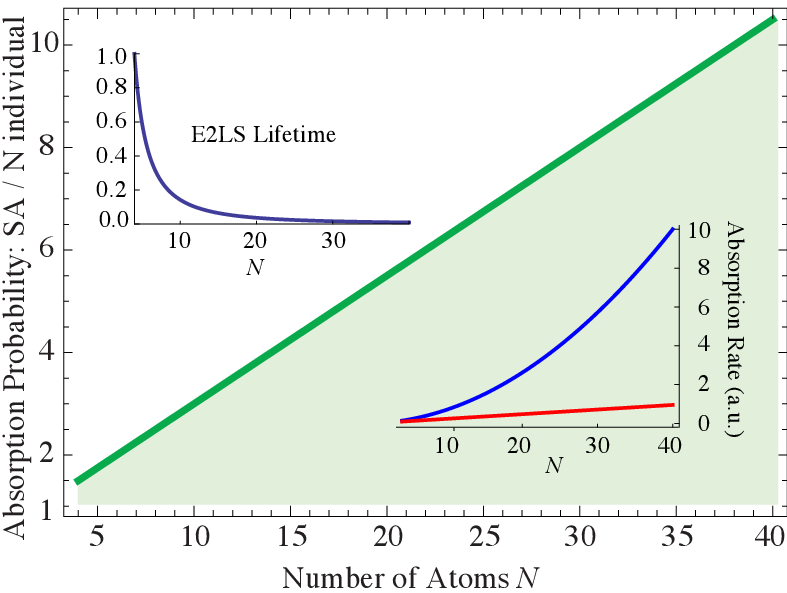}
\caption{\label{fig:new} {\bf Main plot:} probability of absorbing a photon within the lifetime $\Gamma_{\rm loss}^{-1} (N)$ of the superabsorbing E2LS comprising $N$ atoms, compared to that of $N$ individual atoms over the same duration. The relative advantage is linear in $N$ as expected, and the coloured shading indicates the quantum advantage.  {\bf Upper left inset:} lifetime of the E2LS for growing $N$ relative to the four atoms case $\Gamma_{\rm loss}^{-1} (N) / \Gamma_{\rm loss}^{-1} (N=4)$. Note that the decrease in lifetime corresponds to an increasing time resolution of a superabsorbing photon detector: after initialisation the system is receptive to a photon of the requisite frequency only during this time window. {\bf Lower right inset:} absorption rate at the midpoint of the Dicke ladder (blue) and for $N$ individual absorbers (red). The clearly visible $N^2$ scaling that is typical of superradiant pulses also applies to the absorption rate.}
\end{figure}
\begin{figure}
\includegraphics[width=\linewidth]{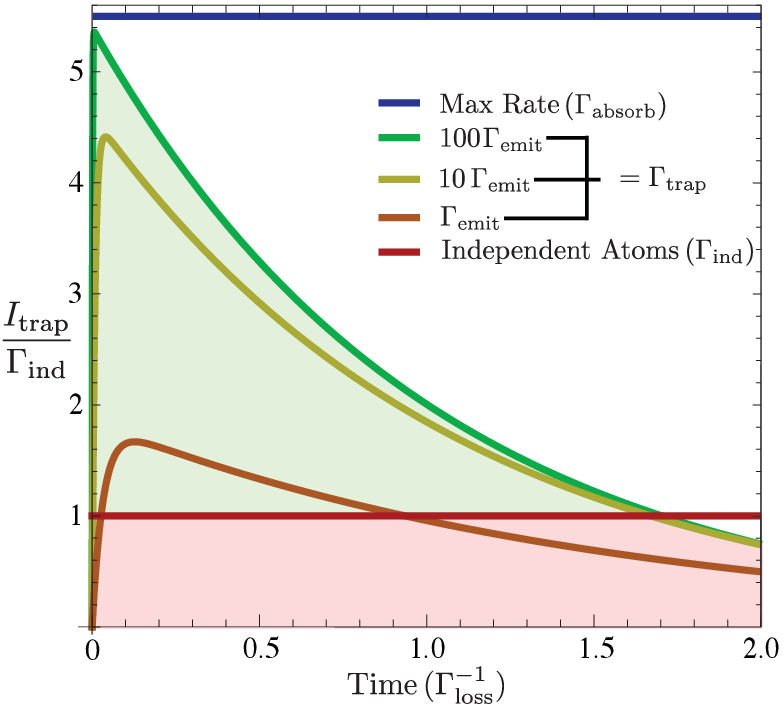}
\caption{\label{fig:3} The transient superabsorption of the effective two level system indicated in Fig.~\ref{fig:1}. The green shading indicates the superabsorption region, the red when the extraction rate is below what could be extracted from uncorrelated atoms; both are for a system of twenty atoms  and mode occupancy $n(\omega_{\rm good})=10$.  The maximum extraction possible from independent atoms ($\Gamma_{\rm ind}= n(\omega_{\rm good}) N \gamma )$ is used for comparison.}
\end{figure}
Our aim is to enhance transition rates at the frequency of the E2LS, which we shall call the `good' frequency $(\omega_{0 \rightarrow -1} \eqdef \omega_{\rm good})$ and suppress those for transitions directly out of the E2LS at the `bad' frequency $(\omega_{-1 \to -2} \eqdef \omega_{\rm bad})$.   The required type of control of the environment is known as reservoir engineering \cite{pechen2006}; in principle we have a choice between tailoring $\kappa(\omega)$, $n(\omega)$, or both. Tailoring the spectral density has the advantage that it can, in theory, completely eliminate the rate of loss from our E2LS when there is no mode of the right frequency present to allow decay. This requires placing the device inside a suitably designed cavity or a photonic bandgap (PBG) crystal with a stop band at $\omega_{\rm bad}$ (see Fig.~\ref{fig:1}c), where the required dimensionality of the PBG depends on the orientation of the optical dipoles. Suppression of emission rates by several orders of magnitude is then achievable with state-of-the-art systems \cite{wang2011, noda2007, leistikow2011}.

Control of $n(\omega)$ is easier to achieve, e.g. by using filtered thermal or pseudothermal \cite{martienssen1964, goodman1985} light. However, this approach has the limitation that even in the optimal control regime, where $n(\omega)=0$ everywhere except in a narrow region around $\omega_{\rm good}$, spontaneous emission will still cause loss from  the E2LS. 

Since both environmental control approaches rely on frequency selectivity, a sufficiently large detuning between adjacent Dicke transitions will be critical for achieving effective containment within the E2LS. In practice the environmental control will never be quite perfect and our system will over long times inevitably evolve away from the E2LS. For example, one may only have control over $n(\omega)$ but not $\kappa(\omega)$, or an imperfect PBG with $\kappa(\omega_{\rm bad}) > 0$, and both cases lead to an exponential decay of E2LS population with the lifetime $\Gamma_{\rm loss}^{-1}$. Dephasing processes will also lead to leakage out of the fully symmetric subspace and thus shorten the effective lifetime of the E2LS. 
However, these imperfections need not dominate the behaviour and destroy the effect. We shall discuss the issue of sustained operation through periodic reinitialisation into the E2LS in Section \ref{sec_reinit}}. 

Let us now consider the properties of the system immediately following initialisation: Fig.~\ref{fig:new} shows the increased photon absorption rate of the superabsorbing E2LS compared to $N$ uncorrelated systems, $\Gamma_{-1 \to 0}  /  N $. Clearly, the probability of absorbing a photon within a given time window (up to the E2LS lifetime) is much higher in the superabsorbing configuration, providing an opportunity for photon dectection with improved sensitivity. The inset of Fig.~\ref{fig:new} shows the lifetime of the E2LS, $\Gamma_{\rm loss}^{-1} = \left( \gamma \Gamma_{-1 \to -2} \kappa(\omega_{\rm bad}) \right)^{-1}$, as a function of $N$, here assumed to be limited by an imperfect PBG with $\kappa(\omega_{\rm bad}) / \kappa(\omega_{\rm good}) = 1/100$. For photon sensing, the reduction of the operational window with increasing $N$ may even be a desirable attribute (offering time resolved detection). Generally, the system we have so far described can function as  sensor as long as the temporary presence of an additional exciton can be registered, for example through continuously monitoring the system's charge state with a quantum point contact \cite{gurvitz1997, goan2011, elzerman2003, petta2004}

\subsection{Trapping}\label{sec_trap}

We have detailed how to create a photon sensor using superabsorption. 
We can also employ the superabsorption phenomenon in the context of energy harvesting if we can meet a further requirement:
a non-radiative channel to extract excitons from the upper of these two levels, turning them into useful work as depicted in the dashed box of Fig.~\ref{fig:1}b. Specifically, we require an irreversible trapping process that extracts only the excitons that are absorbed by the E2LS, and does not extract excitons from levels below the E2LS. Moreover, the trapping process competes with the re-emission of the photons at a rate proportional to $n(\omega_{\rm good}) + 1$, so that ideally it is much faster than that. Note that in this limit saturation is not an issue since absorbed photons are quickly transferred and converted, leaving the system free to absorb the next photon.

The excitons are delocalised across the ring and need to be extracted collectively to preserve the symmetry of the Dicke states. In designing this process we take inspiration from natural light harvesting systems: A `trap' atom is located at the centre of the ring and symmetrically coupled via a resonant hopping interaction to all the other atoms (see Fig.~\ref{fig:schematic}). The corresponding trapping Hamiltonian is
\begin{equation}\label{Trap1}
\hat{H}_{\mathrm{T}} = g (\hat{J}_{+}\hat{\sigma}_{-}^{T}+ \hat{J}_{-}\hat{\sigma}_{+}^{T}) +\omega_{\rm trap}\hat{\sigma}_{+}^{T}\hat{\sigma}_{-}^{T} \, , 
\end{equation}
where the superscript $T$ denotes the trap site, $g$ is the strength of the coupling between the ring and the trap, and the trap's transition frequency $\omega_{\rm trap}$ ideally matches $\omega_{\rm good}$. In this case the interaction is mediated by the electromagnetic field as described in the previous section, but it could have other physical origins depending on the system of interest. Once an exciton has moved to the trap site we assume that it is then removed into the wider environment by a process which irreversibly absorbs its energy. We note that more exotic and potentially far more efficient trapping implementations can be envisioned, such as e.g.~a reservoir of excitons with an effective `Fermi level'  capable of accepting an excitons only above the energy level $E_{-1}$. However, at present our aim is to focus on the simplest system capable of exhibiting enhanced photon energy harvesting by superabsorption. 

The above trapping process is adequately described phenomenologically (see Supplementary Information) as collective exciton extraction from the mid point ($M=0$) by adding the dissipator $D[\hat{L}_{\rm trap}] \rho$ to the righthand side of Eq.~(\ref{MEgen}) with $\hat{L}_{\rm trap} = \sqrt{\Gamma_{\rm trap}} \ket{J, -1} \bra{J,0}$, and where $\Gamma_{\rm trap}$ is the rate of the trapping process. The rate of exciton extraction $I_{\rm trap}$ is then simply given by the population of the trapping level multiplied by the trapping rate: 
\begin{equation}\label{TrapCurrent}
I_{\rm trap}(t)=\Gamma_{\rm trap} \, {\rm{Tr}} \left[ \kb{J,0}{J, 0} \rho(t) \right]  \, .
\end{equation}
Consider an ideal E2LS realised by a PBG completely blocking $\omega_{\rm bad}$, i.e. a vanishing $\Gamma_{\rm loss} \defeq \kappa(\omega_{\rm bad})(n(\omega_{\rm bad})+1)\Gamma_{-1\rightarrow -2}$. Assuming a faster trapping than emission rate, $\Gamma_{\rm trap}\gg \Gamma_{\rm emit} \defeq 
\kappa(\omega_{\rm good})(n(\omega_{\rm good})+1)\Gamma_{0\rightarrow -1}$, our figure of merit $I_{\rm trap}$ matches the absorption rate $\Gamma_{\rm absorb}  \defeq \kappa(\omega_{\rm good})n(\omega_{\rm good})\Gamma_{-1\rightarrow 0}$ for all $t$:
\begin{equation}\label{Trap_Max} 
I_{\rm trap}(t) = I_{\rm max} \approx \Gamma_{\rm absorb} \approx \mu \left( \frac{N}{2} + \frac{N^{2}}{4} \right) \, ,
\end{equation}
where  $\mu =\gamma \kappa(\omega_{\rm good})n(\omega_{\rm good}) $. It is clear from this equation that under these conditions we achieve a superlinear scaling of the exciton current flowing out of the superabsorber. 

The inevitable loss out of the E2LS entails an exponential decay of $I_{\rm trap}(t)$ with the lifetime $\Gamma_{\rm loss}^{-1}$, as shown in Fig.~\ref{fig:3}. The initial net superabsorption rate far exceeds that possible from uncorrelated atoms, however it is only a transient effect and the system needs to be reinitialised periodically to maintain its advantage.  This aspect will be discussed in the next section. 

We have detailed the case where a PBG is used to increase the lifetime of the E2LS. If instead intense filtered thermal light is used to ensure $n(\omega_{\rm good}) \gg 1$, then many absorption-trapping cycles can take place before a spontaneous emission event happens. This set-up would enable quantum enhanced light-based power transmission, where a large number of photons need to be harvested quickly in a confined area.

\subsection{Reinitialisation}\label{sec_reinit}
Reinitialisation could be achieved by exploiting a chirped pulse of laser light to re-excite the system, or through a temporary reversal of the trapping process.  In practice there will be an energy cost associated with reinitialisation but, as we show below, in all but the most severe cases this cost is more than offset by the faster photon to exciton conversion rate during the transient superabsorption periods. 

\begin{figure}
\includegraphics[width=\linewidth]{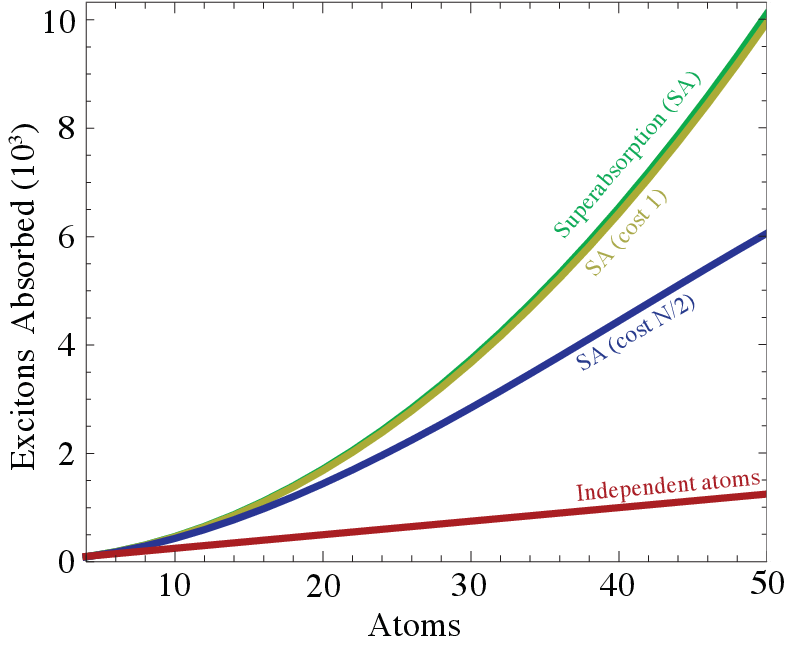}
\caption{\label{fig:4}The total number of excitons absorbed within the common reference time $\Gamma^{-1}_{\rm loss}(N=4)$ as function of the number of atoms $N$. The coloured curves represent the reinitialisation cost models described in the main text, and the red line shows the maximum extracted from independent atoms for comparison. The scaling is superlinear in all coupled atom cases, approximately following the ideal $N^2$ law (green), except for large $N$ in the pessimistic cost model of full reinitialisation (blue). If quantum feedback control enables the replacement of a single exciton as soon as a loss event has happened, then the nearly quadratic scaling persists up to arbitrary numbers of atoms (olive). }

\end{figure}

Perhaps the most elegant way of implementing the reinitialisation step (short of self-initialisation, see below) would make use of quantum feedback control \cite{wiseman10}: The superabsorption enhancement is derived from coherence between states that all possess the same number of excitons. Therefore, the number of excitons could be continually monitored (e.g. by a quantum point contact or by monitoring fluorescence of a probe field tuned to a level or two below the desired manifold) without destroying the desired effect. A suitably designed feedback system could then feed in an excitation only when a loss event had occurred, providing optimal efficiency
\begin{align}\label{Trap} 
 I_{\rm trap}&=\Gamma_{\rm absorb}-\Gamma_{\rm loss} = (\mu-\sigma) \left( \frac{N}{2}+\frac{N^{2}}{4} \right)+2 \sigma \, ,
\end{align}
where $\sigma=\gamma \kappa(\omega_{\rm bad})(1+n(\omega_{\rm bad}))$. Provided $\mu>\sigma$ superabsorption will occur, and for $\sigma=0$ we recover the theoretical maximum of the idealised case in Eq.~(\ref{Trap_Max}).  

A far simpler reinitialisation scheme would only require periodic reinitialisation following a fixed time interval, and does away with the need for feedback control.
In order to account for the relative cost of such reinitialisation, we need to quantify the total number of excitons absorbed in a given time. Let us fix the time at which reinitialisation is performed to the natural lifetime of the E2LS, $\Gamma_{\rm loss}^{-1}$. Integrating the trapping rate $I_{\rm trap}(t)$ over one lifetime and subtracting the reinitialisation cost gives a fair measure of the number of excitons the system has absorbed within the given time. We can then consider the extreme limits of the reinitialization cost, from simply replacing a single lost exciton, to having to replace all of the $N/2$ excitons that make up the superabsorbing state. A larger system requires more frequent reinitialisations, since its loss rate is also enhanced by the system size. However, the bias in favour of absorption created by the environmental control is sufficient to ensure this does not negate the superabsorption process. Fig.~\ref{fig:4} shows how the number of excitons absorbed in a given time scales with the number of atoms, and for all cost models we find a superlinear scaling.

\section{Discussion}

We have shown that the absorptive analogue of quantum superradiance can be engineered in structures with suitably symmetric interactions. We have provided an intuitive explanation of this many-body light-matter effect by introducing an effective two-level system. Despite its simplicity this analytic model can provide highly accurate predictions, as we have validated through the extensive exact numerical calculations that are summarised in the Methods section and presented in the Supplementary Information. As we have already discussed, absorbing light beyond the limits of classical physics raises prospects for at least two new types of technology, and such superabsorption could be realised in a broad range of candidate systems.

The foremost application of the phenomenon may be in the context of optical or microwave sensors, either in future cameras or for scientific instruments. In addition to the obvious merits of being sensitive to low light levels, the frequency specificity of the superabsorber may be a desirable attribute. The small size of the ring structure and collective `antenna array' could lead to high spatial and angular resolution, and the fact that the superabsorber is (re)initialised into its fully receptive state by an excitation pulse allows detection events to be confined into a narrowly defined time window. Note that for sensing applications the cost of (re)initialisation is likely unimportant, and a trapping mechanism is not required if the number of excitons in the system can be monitored differently, e.g. with a quantum point contact.

Light harvesting technologies represent another potential application, and indeed our Fig.~\ref{fig:4} indicates that one can obtain a net increase in the number of exctions absorbed compared to conventional systems even allowing for the energy cost of sustaining the superabsorbing state. The technique would be particularly suited to wireless power transfer using narrowband light, e.g. for remote sensor or biologically implanted devices, where wired electrical power is impractical. For solar light harvesting a given superabsorber can only achieve optimal performance for a specific frequency range; however, one could engineer a range of different systems to jointly cover the solar spectrum. 

There are multiple candidate systems for engineering the above applications. Molecular rings have the advantage of featuring a natural symmetry and intrinsically low levels of disorder. Taking $\Omega=350~\textrm{cm}^{-1}$ as appropriate for a B850 ring \cite{oijen1999} with eight atoms produces transition wavelength shifts exceeding 6~nm, and a wavelength selectivity on the scale of nanometers is readily available with current laser and cavity linewidths. Of course, the dipole alignment of the B850 ring is not optimised for this purpose. However, complex ring structures can be designed and synthesised artificially (for example, porphyrin rings \cite{osullivan2011}) and this route should enable far superior properties. Self-assembly into much larger molecular J or H aggregates with established superradiant properties \cite{mchale2012, saikin2013} may provide further opportunities.
Alternatively, superradiance, long-range interactions, and optical control have been demonstrated in quantum dots~\cite{scheibner2007, unold2005}, and there has been recent progress in synthesising ring like clusters with high spectral and spatial order~\cite{creasey2012}. Further, suppression of the local density of optical states by two orders of magnitude at specific  frequencies has been demonstrated in an appropriate semiconductor photonic crystal environment~\cite{wang2011}. For typical parameters of those systems that would be consistent with the requirements for superabsorption see the Supplementary Information.

To demonstrate the effect of superabsorption (i.e.~sustained confinement into an E2LS with enhanced absorption and emission rates) as an instance of an engineered physical phenomenon, several additional possibilities present themselves. For example, circuit QED experiments possess long coherence times and have already demonstrated sub- and superradiant effects \cite{fink2009, filipp2011} as well as tuneable cross Lamb shifts \cite{lalumiere2013}, and recent 3D structures \cite{paik2011} provide further flexibility. Bose Einstein Condensates offer similar properties but with much larger numbers of atoms \cite{colombe2007, nagy2010}. Dissipative Dicke model studies with nonlinear atom-photon interaction can enable a steady-state at the mid-point of the Dicke ladder ($M=0$)~\cite{dimer2007,grimsmo2013}, which may provide a route to self-initialising superabsorbing systems.

\section{Methods}

\subsection{Collective Master Equation }

The master equation (\ref{MEgen}) is an $N$ atom generalisation of the standard quantum optical master equation, we give the full derivation in the Supplementary Information document. In particular, it assumes that all $N$ atoms are spatially indistinguishable due to occupying a volume with linear dimensions much smaller than the relevant wavelength of light. In addition, interactions between atoms must respect certain symmetry requirements to only shift the Dicke states to first order [e.g.~as is exemplified by Eq.~(\ref{eq:energy_shift})]. However, as we also discuss in the SI -- and verify with numerical calculations -- superlinear scaling of the absorption rate with the number of atoms remains possible beyond a first order perturbative treatment of suitably symmetrical interactions.

\subsection{Numerical Calculations}

The Effective Two Level System (E2LS) model reduces the complexity of the problem and makes it analytically tractable. In order to verify this approach we have compared it to two different independent numerical models. In Fig. S5 in the Supplementary Information shows excellent agreement between the E2LS model and Monte Carlo simulations of the master equation (\ref{MEgen}). In Fig S2, we extend the model the model further by incorporating an explicit trap site and allow coherent transfer from the ring to the trap, as described in Section \ref{sec_trap}, showing that superabsorption is still realised in that case and that the E2LS model still provides a good description of the behaviour. This model uses a generalised master equation solved  numerically. 

\subsection{Imperfections}
Any real physical system used to demonstrate superabsorption, or indeed superradiance, will have imperfections  such as slightly varying frequencies for each atom, or a deviation away from perfect ring symmetry. In essence all such imperfections in superradiance are alike; they diminish the collective effect because they lead to the emission of distinguishable photons.  It might therefore be a concern that these collective effects could only be realised in the ideal case.  However, superradiant effects of molecular aggregates with a spatial extent smaller than the wavelength of light are known to possess a certain degree of robustness against inhomogenous broadening \cite{spano1989}, dephasing processes \cite{grad1988}, and exciton phonon coupling \cite{spano1991}. This is because the increased transition rates produced by superradiance serve to counterbalance the effect of disorder:  the faster rate broadens the natural linewidth of the transitions, effectively masking the distinguishably introduced by the disorder.  Intuitively, we expect a superabsorption advantage to be achievable whenever an imperfect system is still capable of displaying superradiant behaviour (of course with the additional requirement that the energy shifts of adjacent decay process are resolvable). In the SI,  we model realistic imperfections by considering static energy disorder and show that superabsorption can still be realised in the presence of disorder.

\section{Acknowledgments}
This work was supported by the EPSRC, the Leverhulme Trust, the National Research Foundation and Ministry of Education, Singapore, and ARC grant CE110001013. BWL thanks the Royal Society for a University Research Fellowship.

\section{Author contributions}
All authors designed the protocol, analysed the results, and discussed the manuscript. KDBH, BWL, and EMG performed the calculations. KDBH, SCB, BWL, and EMG wrote the manuscript. 

\section{References}

\bibliographystyle{apsrev}

\end{document}


\title{Supplementary Information: Superabsorption of light via quantum engineering}
\author{K. D. B. Higgins}
\affiliation{Department of Materials, Oxford University, Oxford OX1 3PH, United Kingdom}
\author{S. C. Benjamin}
\affiliation{Department of Materials, Oxford University, Oxford OX1 3PH, United Kingdom}
\affiliation{Centre for Quantum Technologies, National University of Singapore, 3 Science Drive 2, Singapore 117543}
\author{T. M. Stace}
\affiliation{Centre for Engineered Quantum Systems, School of Mathematics and Physics, The University of Queensland,
St Lucia, Queensland 4072, Australia}
\author{G. J. Milburn}
\affiliation{Centre for Engineered Quantum Systems, School of Mathematics and Physics, The University of Queensland,
St Lucia, Queensland 4072, Australia}
\author{B. W. Lovett}
\affiliation{SUPA, Department of Physics, Heriot Watt University, Edinburgh EH14 4AS, United Kingdom}
\author{E. M. Gauger}
\affiliation{Centre for Quantum Technologies, National University of Singapore, 3 Science Drive 2, Singapore 117543}
\affiliation{Department of Materials, Oxford University, Oxford OX1 3PH, United Kingdom}

\maketitle

\section{Master Equation Derivation}\label{ME_deriv}

In the following we sketch the derivation of the master equation (7) from the main Article. Following the general procedure of Ref.~\cite{gross1982}, we generalise from a vacuum environment to one with a population distribution and structured spectral density. This allows for the presence of the superabsorption term and introduces some additional complexities. 

We consider the interaction picture with respect to $\hat{H}_S$ [Eq.~(1) of the Article] and the free Hamiltonian of the electromagnetic field. After performing the standard Born-Markov approximation and tracing over the environment, the starting point for our derivation is \cite{breuer02} ($\hbar =1$):
\begin{equation}\label{master4}
\frac{d}{dt}\tilde{\rho}_{\cal{S}}(t)=-i \int_0^\infty{dt'\text{Tr}_{\cal{E}}[\tilde{H}_{L}(t), [\tilde{H}_{L}(t-t'), \tilde{\rho}_{\cal{S}}(t) \otimes \tilde{\rho}_{\cal{E}}]} + \mathrm{h.c.}\, ,
\end{equation}
where $\tilde{\rho}_{\cal{S}}(t)$ is the reduced interaction picture density matrix and $\tilde{H}_{L}(t)$ denotes the interaction picture representation of the system-light-interaction Hamiltonian $\hat{H}_{L}= -\sum_{i=1}^{N}  \hat{\sigma}^{i}_{-} \, \vec{d} \cdot \hat{E}(\vec{r}_{i}) +  \vec{\sigma}^{i}_{+} \vec{d}^{*} \cdot \vec{E}(\vec{r}_{i}) \, ,$
 [c.f. Eq.~(3) of the main text]. Here, $\vec{d}$ is  the atomic dipole vector and the electric field operator is given by
  \begin{equation}\label{E_op}
\hat{E}(\vec{r}_{i})=i \sum_{i=1}^{N}\sum_{\vec{k},\lambda}\sqrt{\frac{2 \omega_{k}}{V}}\vec{e}_{\lambda}(\vec{k}) \left( \hat{b}_{\lambda}(\vec{k})e^{i\vec{k} \cdot \vec{r}_{i}}-\hat{b}^{\dagger}_{\lambda}(\vec{k})e^{-i\vec{k} \cdot \vec{r}_{i}} \right),
\end{equation} 
where $\vec{e}_{\lambda}(\vec{k})$ and $\hat{b}^{(\dagger)}_{\lambda}(\vec{k})$ are the polarisation vector of the field and its annihilation (creation) operator, respectively. The system dynamics is then generically determined by the following master equation \cite{breuer02}
 \begin{equation} \label{GenericFullME}
\frac{d}{dt} \rho_{S}(t) = - i [\hat{H}_S+\hat{H}_{I}, \rho_{S}(t)]  + \sum_{\omega} \sum_{i, j} \left[ \Gamma_{i, j}(\omega)(\hat{A}_{j}\rho_{S}(t) \hat{A}^{\dagger}_{i}- \hat{A}^{\dagger}_{i} \hat{A}_{j}\rho_{S}(t)) + \mathrm{h.c.} \right],
 \end{equation}
 where H.c.~denotes the Hermitian conjugate. The $\hat{A}_{i}$ are the Lindblad operators given by $\hat{\sigma}_-^i$ and $\hat{\sigma}_+^i$, and $\Gamma_{i,j}=\int_0^\infty ds \, e^{i\omega s}\langle \vec{d^{*}} \cdot \hat{E}(\vec{r}_{i},s)  \vec{d} \cdot  \hat{E}(\vec{r}_{j},0) \rangle$ is the spectral correlation tensor, which will be calculated in the following. We start by considering the expression
\begin{equation}\label{SCT_op1}
\langle  \vec{d^{*}} \cdot  \hat{E}(\vec{r}_{i},s)   \vec{d} \cdot \hat{E}(\vec{r}_{j},0) \rangle= Tr_{E}[  \vec{d^{*}} \cdot  \hat{E}(\vec{r}_{i},s)   \vec{d} \cdot \hat{E}(\vec{r}_{j},0)\rho_{E}],
\end{equation} 
where $\rho_{E}$ denotes the thermal state of the environment, though allowing filtered thermal light later will not change the form of the result \cite{pechen2011, accardi2002}. Generally, for a thermalised environment it is well known that \cite{breuer02}
\begin{align} \label{SCT_op2}
&\langle \hat{b}_{\lambda}(\vec{k}) \hat{b}_{\lambda'}(\vec{k'})\rangle= \langle( \hat{b}^{\dagger}_{\lambda}(\vec{k}) \hat{b}^{\dagger}_{\lambda'}(\vec{k'})\rangle=0, \\
&\langle \hat{b}_{\lambda}(\vec{k}) \hat{b}^{\dagger}_{\lambda'}(\vec{k'})\rangle=\delta_{\vec{k}\vec{k}'}\delta_{\hat{\lambda}\hat{\lambda}'}=(1+n(\omega_{k})), \\
&\langle \hat{b}^{\dagger}_{\lambda}(\vec{k}) \hat{b}_{\lambda'}(\vec{k'})\rangle=\delta_{\vec{k}\vec{k}'}\delta_{\hat{\lambda}\hat{\lambda}'}=n(\omega_{k}).
\end{align} 
Using these, the spectral correlation tensor can written as:
\begin{equation}\label{SCT_op3}
\Gamma_{i,j}=\frac{2 \pi }{V}\sum_{\vec{k},\lambda}( \vec{d} \cdot \vec{e}_{\lambda}(\vec{k}))^{2}\omega_{k}\bigg((1+n(\omega_{k})) e^{i\vec{k}\cdot \vec{r}_{ij}}
\int_0^\infty ds \, e^{-i(\omega_{k}-\omega) s}+e^{-i\vec{k}\cdot \vec{r}_{ij}}n(\omega_{k}) \int_0^\infty ds \, e^{i(\omega_{k}+\omega) s}\bigg),
\end{equation}
where $\vec{r}_{i,j}$ is the vector connecting atoms $i$ and $j$. Converting the sum over $\vec{k}$ to an integral ($\omega_k = c \vert \vec{k} \vert$) yields
\begin{align}\label{sum2int}
\frac{1}{V}\sum_{\vec{k}}\rightarrow \frac{1}{(2\pi)^3 c^3}\int_{0}^{\infty}d\omega_k \, \kappa(\omega_k)\omega_k^2\int d\Omega  \, ,
\end{align}
where $\kappa(\omega)$ is the spectral density given by the density of states weighted by the coupling strength, $\kappa(\omega)= \sum_{k}\abs{g_{k}}^2\delta(\omega-\omega_{k})=\chi({\omega})\abs{g(\omega)}^{2}$.
The angular part of the integration gives a diffraction type function:
\begin{align}
F(\omega \, \vec{r}_{ij})=\frac{8\pi}{3}\left(j_{0}(\omega \, \vert \vec{r}_{ij} \vert)+\frac{1}{2}\left(3\cos^{2}(\theta_{\vec{d} \, \vec{r}_{ij}})-1\right)j_{2}(\omega \, \vert \vec{r}_{ij} \vert)\right)  \, ,
\end{align}
where $j_{n}(x)$ is the $n^{th}$ spherical Bessel function and the angle $\theta_{\vec{d} \, \vec{r}_{ij}}$ between the atomic dipoles and pairwise connection vectors is 
\begin{equation}
\cos^{2}{\theta_{\vec{d} \, \vec{r}_{ij}}}=\frac{\vert \vec{d}\cdot \vec{r}_{ij}^{\,2} \vert}{\vert\vec{d}^2 \vert \vert{\vec{r}_{ij}}^{\,2}\vert} \,.
\end{equation}
Considering the geometry in Fig.~1 of the Article, we assume that all dipoles are parallel, and perpendicular to the plane defined by the ring. In this case, and for only  nearest neighbour interactions, $\theta_{\vec{d} \, \vec{r}_{ij}}$ is  independent of $i$ and $j$. 
Thus Eq.~(\ref{SCT_op3}) becomes
\begin{equation}\label{SCT_op4}
\Gamma_{i,j}=\frac{\abs{d}^2}{(2\pi)^2c^3}\int_{0}^{\infty}d\omega_{k}\, \kappa(\omega_{k})\omega_{k}^3 \, F(\omega_{k} r_{ij})
\bigg((1+n(\omega_{k})) \int_0^\infty ds e^{-i(\omega_{k}-\omega) s}+n(\omega_{k}) \int_0^\infty ds e^{i(\omega_{k}+\omega) s}\bigg)  \, ,
\end{equation}
which we separate into its real and imaginary parts $\Gamma_{i,j}= \frac{1}{2} \gamma_{i,j}(\omega)+i S(\omega)$ with the help of the identity
\begin{equation}\label{SCT_indent}
\int_0^\infty ds \, e^{\pm i \epsilon s}= \pi \delta(\epsilon) \pm iP\frac{1}{\epsilon}.
\end{equation}
The real terms $\gamma_{i,j}(\omega)$ derive from the delta functions and give rise to the dissipative dynamics, i.e. optical transitions in this case. In the remaining term, $F(\omega_{k} \, r_{ij})$ is evaluated at $\omega_{k}= \pm \omega$. We are working in the small sample limit, where the wavelength of light is far longer than the size of our nanostructure ($\omega \, r_{ij} \approx 0$), and so $F(\omega \, r_{ij})\approx 8 \pi / 3 $. Hence $\gamma_{i,j}$ is independent of the atomic indices to a good approximation:
\begin{equation}\label{SCT_op}
 \gamma_{i,j}(\omega)\approx  \gamma(\omega) = \frac{4 \omega^3 \abs{d}^2}{3 c^3}\kappa(\omega)(1+n(\omega))\, 
\end{equation}
The Planck distribution has the property that $n(-\omega)=-(1+n(\omega))$. Thus we can combine the terms arising from $\delta(\omega_{k} \pm \omega)$ and only run the sum over positive values.  The second term on the righthand side of Eq.~(\ref{GenericFullME}) thus becomes
\begin{align}
\sum_{\omega>0}\sum_{i, j} \frac{4 \omega^3 \abs{d}^2}{3 c^3}\kappa(\omega) \bigg((1+n(\omega))(\sigma^{-}_{j} \rho \sigma^{+}_{i} -\frac{1}{2} \left\{ \sigma^{+}_{i}\sigma^{-}_{j}  , \rho \right\})
						    		    +n(\omega)(\sigma^{+}_{j} \rho \sigma^{-}_{i} -\frac{1}{2} \left\{ \sigma^{-}_{i}\sigma^{+}_{j}  , \rho \right\}) \bigg)  \, .
\end{align}
By simply assuming all transitions have the same frequency splitting ($\omega=\omega_{A})$,  a vacuum environment state $n(\omega) = 0$ and switching to the collective operators to express the sums $\hat{J}_{-} = \sum_{i} \hat{\sigma}^{-}_{i}$, we reproduce the ordinary superradiance master equation. \\

We now turn to the imaginary part $S(\omega)$ of the spectral correlation tensor, which will be responsible for providing the detuning between different transitions; this is given by:
\begin{align}
S(\omega) =  \frac{\abs{d}^2}{(2\pi)^2c^3}P\int_{0}^{\infty}d\omega_{k}\, \kappa(\omega_{k}) \omega_{k}^3 \, F(\omega_{k} \, r_{ij})
\bigg(\frac{1+n(\omega_{k})}{\omega-\omega_{k}}+ \frac{n(\omega_{k})}{\omega_{k}+\omega}\bigg)  \, .
\end{align}
The $i=j$ terms, for which $F(0)= 8 \pi / 3$, correspond to the ordinary Lamb shift of individual atom transitions; these can be accounted for by a renormalisation of the bare atomic frequency $\omega_A$.
By contrast, the $i\neq j$ terms correspond to the dipole-dipole interaction induced by the EM field. Evaluating this integral requires us to choose a specific form for the spectral density $\kappa(\omega)$. Here we consider two cases: first a flat spectral density, and second one that features a `stop band'  in the spectrum, blocking the `bad' transition at frequency $\omega_{\rm bad}$. We begin with the former case. We first separate out the term that is independent of $n(\omega)$, the Lamb shift $S_{L}(\omega)$, and evaluate it. In the small sample limit taken  ($\omega r_{ij} \ll 1$) we find
 \begin{align}\label{DDinteraction}
S_{L}(\omega)= \frac{d^2}{4\pi \epsilon_{0} r^{3}_{ij}} \left[1-3 \cos^2(\theta_{dr})\right] \, .
\end{align}
After separating out the Lamb shift, we are left with the divergent integral corresponding to the Stark shift: 
\begin{align}\label{Stark1}
S_{s}(\omega) =  \frac{\abs{d}^2}{(2\pi)^2c^3}P\int_{0}^{\infty}d\omega_{k}\,  \kappa(\omega_{k}) \omega_{k}^3 \, F(\omega_{k} r_{ij})
\bigg(\frac{n(\omega_{k})}{\omega-\omega_{k}}+ \frac{n(\omega_{k})}{\omega+\omega_{k}}\bigg)  \, .
\end{align}
This very seldom evaluated in the literature and is usually assumed negligible. In this work we are primarily concerned with controlling $n(\omega)$ so that it is only significant for one mode, which has frequency $\omega_{g}$. In this case we set $n(\omega_{k})=\delta(\omega_{k}-\omega_{g})$ and take the small sample limit:
\begin{equation}\label{Stark2}
S_{s}(\omega) = \lim_{r \omega_{g} \to 0} \frac{4 \pi  \omega \left(\sin(r \omega_{g}) \left(\cos (2 \theta_{dr}) \left(3-r^2 \omega_{g}^2\right)+r^2 \omega_{g}^2+1\right)-r \omega_{g} (3 \cos(2 \theta_{dr})+1) \cos(r \omega_{g})\right)}{r^3 (\omega-\omega_{g}) (\omega+\omega_{L})}=0  \, .
\end{equation}
Hence we can neglect the Stark shift and only retain the Lamb shift. Returning to the other spectral density we consider in the paper, that with the stop band, we can express the stop band with the following, simplistic spectral destiny: 
\begin{align}\label{Gap1}
\kappa(\omega)= 1-T(\omega_{b},\sigma)  \, ,
\end{align}
where $T(\omega_{b},\sigma)$ is the `top hat' function centred on $\omega_{b}$ with a width $\sigma$. The factor of one produces the same result as for the flat spectral density $S_{s}(\omega)$. The top hat handles the effect of the gap $S_{gap}(\omega)$:
\begin{align}\label{Gap2}
S(\omega)= S_{s}(\omega)-S_{gap}(\omega)  \, .
\end{align}
The top hat has the effect of confining the integral to a window around $\omega_{b}$: 
\begin{align}
S_{gap}(\omega) =  \frac{\abs{d}^2}{(2\pi)^2c^3}P\int_{\omega_{b}-\sigma}^{\omega_{b}+\sigma}d\omega_{k}\, \omega_{k}^3 \, F(\omega_{k} r_{ij})
\bigg(\frac{1+n(\omega_{k})}{\omega-\omega_{k}}+ \frac{n(\omega_{k})}{\omega_{k}+\omega}\bigg) \, .
\end{align}
This  can be evaluated to yield a lengthly, but straightforward expression. For an ideal gap ($ \sigma \to 0$), $S_{gap}(\omega)=0$ and hence can be neglected.  Collating these results, we are left with a familiar expression for the strength of the interaction between two dipoles (\ref{DDinteraction}) multiplied by a hopping term introduced via the EM field:
\begin{equation}\label{Hopping}
\hat{H}_{I}= S_{L}(\omega) \sum_{i \neq j}^{N} \left( \hat{\sigma}_{+}^{i} \hat{\sigma}_{-}^{j}+ \hat{\sigma}_{-}^{i} \hat{\sigma}_{+}^{j} \right) \, .
\end{equation}
In the paper we consider the nearest neighbour limit of this expression, although this is not a necessity (see below). We know that the EM field operators interact with the system collectively, causing them to explore the ladder of Dicke states defined by:
\begin{equation}\label{DickeState}
\ket{J,M}= \sqrt{\frac{(J+M)!}{N! (J-M)!}}\hat{J}_{-}^{(J-M)} \ket{ee....e}  \, .
\end{equation}
The highly symmetric geometry of our system means that the hopping interaction (\ref{Hopping}) does not cause mixing of Dicke levels, but only shifts their energies (see Article). Thus an effective Hamiltonian for the subspace consisting of only the fully symmetric states of the Dicke ladder can be written as :
\begin{equation}\label{HamiltonianD}
\hat{H}_{S}+\hat{H}_{I} = \frac{E_{M}}{2} \sum^{J}_{M=-J} \kb{J,M}{J,M}  \, ,
\end{equation}
where $E_{M}$ is the energy of the $\ket{J,M}$ state now including the shift defined in Eq.~(10) of the Article. Instead of having a generic ladder operator $\hat{J}_{\pm}$ that moves any state $\ket{J,M} \to \ket{J,M \pm1}$, with emission or absorption at $\omega_{A}$, we now have to break up this operator, because the Dicke transitions are no longer degenerate in energy. The generic ladder operators are thus replaced by a sum over operators, which take us between specific Dicke states, sampling the spectral density at the requisite frequency $\omega_{\beta}$:
\begin{equation}\label{Lindblad}
 \hat{L}_{M}=\kb{J,M-1}{J,M} \, , 
\end{equation}
which yields the result in the main paper Eq.~(7): 
\begin{equation}\label{MEgen_Apen}
\dot{\rho}  = - i [\hat{H}_{S}+\hat{H}_{I}, \rho]  - \gamma \sum_{\beta} \kappa(\omega_{\beta}) \left( (n(\omega_{\beta})+1) D[\hat{L}_{\beta}]\rho + n(\omega_{\beta}) D[\hat{L}^{\dagger}_{\beta}]\rho \right) \, .  
\end{equation}

\section{Resolution of Frequency Shifts}

The enhanced absorption and emission rate in the middle of the Dicke ladder implies an increased lifetime broadening. One might thus worry whether the detunings obtained courtesy of the hopping interaction are then still sufficient to completely resolve adjacent transitions. A simple analysis shows that this is indeed the case \cite{gross1982}, and the natural width ($N^{2} \gamma$ around $M=0$) remains smaller than the shift $\delta_{\omega}$ provided the wavelength of the light is much greater than the size of the system. 
To see this, we shall assume that the transition is indeed well resolved, $N^{2} \gamma <  \delta_{\omega}$, and show that is essentially equivalent to the `small sample'  condition, $r \ll \lambda$,  which underlies the phenomenon of superradiance in the first place.

Using the definition of $\gamma$ from as the free atom decay rate (see Article), the greatest broadening and smallest energy shift at $M=0$ are, respectively, 
\begin{align} 
\label{eq:LHS}
N^{2} \gamma & = \frac{ 8 N^{2} \pi^2 d^{2}}{ 3 \epsilon_{0} \hbar \lambda^{3}} \, , \\
\label{eq:RHS}
\delta_{\omega} & = 4 \frac{\Omega}{N-1} \approx \frac{d^2}{4\pi \hbar \epsilon_{0} r^{3}} \, .
\end{align}
Substituting Eqs.~(\ref{eq:LHS}) and (\ref{eq:RHS}), the inequality $N^{2} \gamma <  \delta_{\omega}$ becomes
\begin{align}
 \frac{ 2 N^{3} \pi^2 d^{2}}{ 3 \epsilon_{0} \hbar \lambda^{3}} <  \frac{d^2}{4\pi \hbar \epsilon_{0} r^{3}} \quad \Longleftrightarrow \quad
2 N \pi r  < \lambda \,,
\end{align}
where the righthand side follows after cancellation of several variables followed by taking the qubic root. This is equivalent to $r \ll \lambda$, up to moderate numerical factor (when $N$ is not too large), accounted for by relaxing `$\ll$' to `$<$'. For the present discussion, 
$r$ is understood to be the nearest neighbour distance, having assumed energy shifts appropriate for only nearest neighbour interactions (for other interaction models, the detunings would be larger).
We note that distinct shifted lines have also already been observed --- and resolved --- experimentally \cite{wang1995a}.

\section{Interactions beyond the nearest neighbour limit}

In the main text we assumed only nearest neighbour interactions are significant.  For a symmetric ring geometry relaxing this condition leads to the same qualitative behaviour, but results in slightly larger detunings between adjacent transitions in the Dicke ladder. First, let us consider the opposite limit to the nearest neighbour case and allow all pairwise interactions with equal strength: 
\begin{equation}\label{eq:BNNShift}
\bra{J,M} \hat{H}_{H}\ket{J,M}= \bra{J,M} \Omega \sum_{i \neq j} (\sigma_{i}^{+} \sigma_{j}^{-}+\sigma_{i}^{-} \sigma_{j}^{+})\ket{J,M} \, ,
\end{equation}
\begin{equation}\label{eq:BNNShift2}
\delta E_{m}=  \Omega \bra{J,M} \sum_{i \neq j} (\sigma_{i}^{+} \sigma_{j}^{-}+\sigma_{i}^{-} \sigma_{j}^{+})\ket{J,M} \, ,
\end{equation}
which can be rewritten using the collective operators as follows 
\begin{equation}\label{eq:BNNShift3}
\delta E_{m}=  \Omega \bra{J,M} J_{+}J_{-} +J_{-}J_{+}-\sum_{i} (\sigma_{i}^{+} \sigma_{i}^{-}+\sigma_{i}^{-} \sigma_{i}^{+})\ket{J,M} \, .
\end{equation}
The final two terms are added to remove the $i=j$ terms implicit in the $J_{+}J_{-}$ terms, which count the number of excited and unexcited atoms, respectively. Hence,
\begin{equation}
\delta E_{m}=  \Omega( \bra{J,M} J_{+}J_{-} +J_{-}J_{+}\ket{J,M}-2J) \, .
\end{equation}
The remaining two terms are easily calculated from Eq.~(4) of the Article, yielding:   
\begin{equation}
\delta E_{m}=  2\Omega(J^{2}-M^{2}) \, .
\end{equation}
Thus the energy shifts are the same as in the nearest neighbour case Eq.~(10) but lack the factor$(J-1/2)^{-1}$. Therefore,  unlike in the nearest neighbour limit,  increasing the number of atoms does not reduce the size of the frequency shift, which could help in blocking the transition at $\omega_{bad}$ and ensuring frequency selectivity of a trapping mechanism. 

The actual ring geometry with all pairwise dipole interactions included will fall somewhere in between these two limits, depending on ring size. The operators involved in the interaction remain the same, but their weights are altered as the size of the ring changes. The symmetry of the ring dictates that each atom will be subject to the same set of interactions with the rest of the ring.  The condition of interchangeability of atoms is thus met regardless of the specific interaction model (i.e.~nearest neighbour, next nearest neighbour etc.). For all cases, the hopping interaction only causes shifts of a variable size between the two limits we have discussed above; the size of the shifts given a particular ring size and interaction model is readily obtained numerically.

\section{The Implicit interaction Approach}

In the main paper, the hopping interaction emerged from the derivation of the master equation, as a direct consequence of embedding the absorbers into the photon field environment. Alternatively, we could add an interaction to the Hamiltonian, which may either be mediated by virtual photon exchange or have some other physical origin. The initial Hamiltonian then reads:
\begin{equation}\label{HamiltonianS}
\hat{H}_{S} =  \omega_{A} \sum_{m=1}^{N} \hat{\sigma}^{m}_{+}\hat{\sigma}^{m}_{-}+  \Omega \sum_{i, j}^{N}(\hat{\sigma}_{+}^{i} \hat{\sigma}_{-}^{j}+ \hat{\sigma}_{-}^{j} \hat{\sigma}_{+}^{i}) \, .
\end{equation}
Such a Hamiltonian is a diagonalised using the Jordan-Wigner transformation  \cite{benedict1996}.  For example, a four atom system has eigenvalues:  
\begin{equation}\label{eq:eigvals}
\left\{0, \omega_{A}-2 \Omega ,2 \omega_{A}-2 \sqrt{2} \Omega ,3 \omega_{A} -2 \Omega,4 \omega_{A} \right\} \, ,
\end{equation}

resulting in the following transition frequencies:
\begin{equation}\label{eq:freq}
\omega=\left\{ \omega_{A} -2\Omega, \omega_{A}-2 \Omega \left(1+\sqrt{2}\right) , \omega_{A}+2\Omega \left(-1+\sqrt{2}\right), \omega_{A}+2 \Omega \right\} \,.
\end{equation}
These differ slightly from those derived using the approach in the main paper. Crucially, however, the degeneracy of the transition frequencies is broken in a similar way as before, so that the discussion of environmental control for confining the dynamics to a specific transition remains valid. For small length scale linear systems it has been noted that superradiance dynamics are not significantly altered \cite{benedict1996}, when compared to the traditional field mediated interaction approach \cite{gross1982}, used in the main text.

\begin{figure}
\includegraphics[scale=1.1]{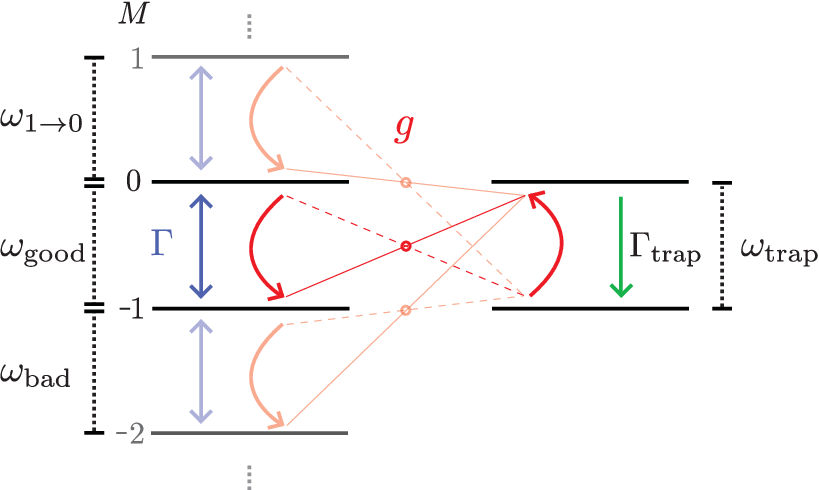}
\caption{\label{fig:5} A trap site is connected to the Dicke ladder. The trap's transition frequency ideally matches that of the `good' transition $\omega_{\rm trap} \approx \omega_{\rm good}$, and it is coupled to the Dicke transitions via a flip-flop interaction of strength $g$. This gives rise to `see-saw' like oscillations between Dicke and trap transitions, but only the desired transition is resonant, all others are detuned and thus suppressed. To ensure that excitons hopping to the trap site are irreversibly removed instead of `see-sawing' back and forth indefinitely, the trap is incoherently emptied at a rate $\Gamma_{\rm trap}$. As long as $\Gamma_{\rm trap}$ is not so large that the trap transition experiences significant lifetime broadening approaching $\vert \omega_{\rm good} - \omega_{\rm bad} \vert$, and also assuming $\vert \omega_{\rm good} - \omega_{\rm bad} \vert > g$,  the exciton extraction from the `bad' transition in particular can be suppressed. The blue double-headed arrows indicate the (enhanced) optical emission and absorption processes from the main paper.}
\end{figure}

\section{Trapping}\label{sec_trap}

In the main text we describe the trapping process using a Lindblad operator, which takes the system from the state $\ket{J,0} \to \ket{J,-1}$ by irreversibly removing one exciton from the system. Numerous microscopic mechanisms can be conceived of that would produce this effect; all will have to involve a collective coupling of the atoms of the ring, followed by a process (through coupling to a wider external environment) which prevents the return of exciton to the primary system, or at least renders it very unlikely. 

Here, we give an account of the simplest scenario one can envision: all atoms of the ring are  coupled to a `trap' atom at the centre, whose role is to first localise the energy and then irreversibly remove it. This simple trapping model is schematically depicted in Fig. \ref{fig:5}: The trap atom possesses a transition frequency $\omega_{\rm trap} \approx \omega_{\rm good}$ and is coupled to the ring by a field mediated hopping interaction of strength $g$ (i.e.~the same type of interaction which couples the ring's atoms to each other). The Hamiltonian for such a set up is:

\begin{equation}\label{H_trap}
\hat{H}_{\mathrm{T}} = g (\hat{J}_{+}\hat{\sigma}_{-}^{T}+ \hat{J}_{-}\hat{\sigma}_{+}^{T}) +\omega_{\rm trap}\hat{\sigma}_{+}^{T}\hat{\sigma}_{-}^{T} \, , 
\end{equation}
where the superscript $T$ denotes the trap site, $g$ is the strength of the coupling between the ring and the trap, and the trap's transition frequency $\omega_{\rm trap}$, which ideally matches $\omega_{\rm good}$. An exciton `hopping' onto the trap site is subjected to an irreversible decay with rate $\Gamma_{\rm trap}$, e.g.~by being linked to a chain of exciton sites acting as a wire or lead. In natural light harvesting systems the trap would be the reaction centre and the decay a photochemical process. 

When the trapping rate $\Gamma_{\rm trap}$ is sufficiently fast only negligible population exists in the 2LS forming the trap, hence its effect can, to a good approximation, be considered that of a Lindblad operator acting on the main system. For a slower rate $\Gamma_{\rm trap}$ a decaying Rabi oscillation may take place, moving the exciton back and forth between trap and ring. However, the presence or absence of these oscillations does not significantly affect the superabsorption process. 

This simplistic trapping model introduces an extra contribution to the rate of loss from the E2LS: the finite lifetime $1/\Gamma_{\rm trap}$ means the energy of the trap is not perfectly sharply defined, reducing the trap's frequency selectivity. Thus, it will occasionally also accept energy from the `bad' transition $\omega_{\rm bad}$, which increases the effective loss rate $\Gamma_{\rm loss}$. However, one can optimise the parameters  ($\Gamma_{\rm trap} \, , g, \, \omega_{\rm{trap}}$) to minimise this undesirable side effect, whilst still meeting the condition that $\Gamma_{\rm trap} > \Gamma_{\rm emit}$. With the addition of the trapping terms (\ref{H_trap}) the master equation is numerically solved using the method described in Section \ref{sec:num_res}.  We compare this against the analytical expression for the trapping rate that can be derived for the E2LS model described in the main paper: 

\begin{figure}
\includegraphics[scale=0.62]{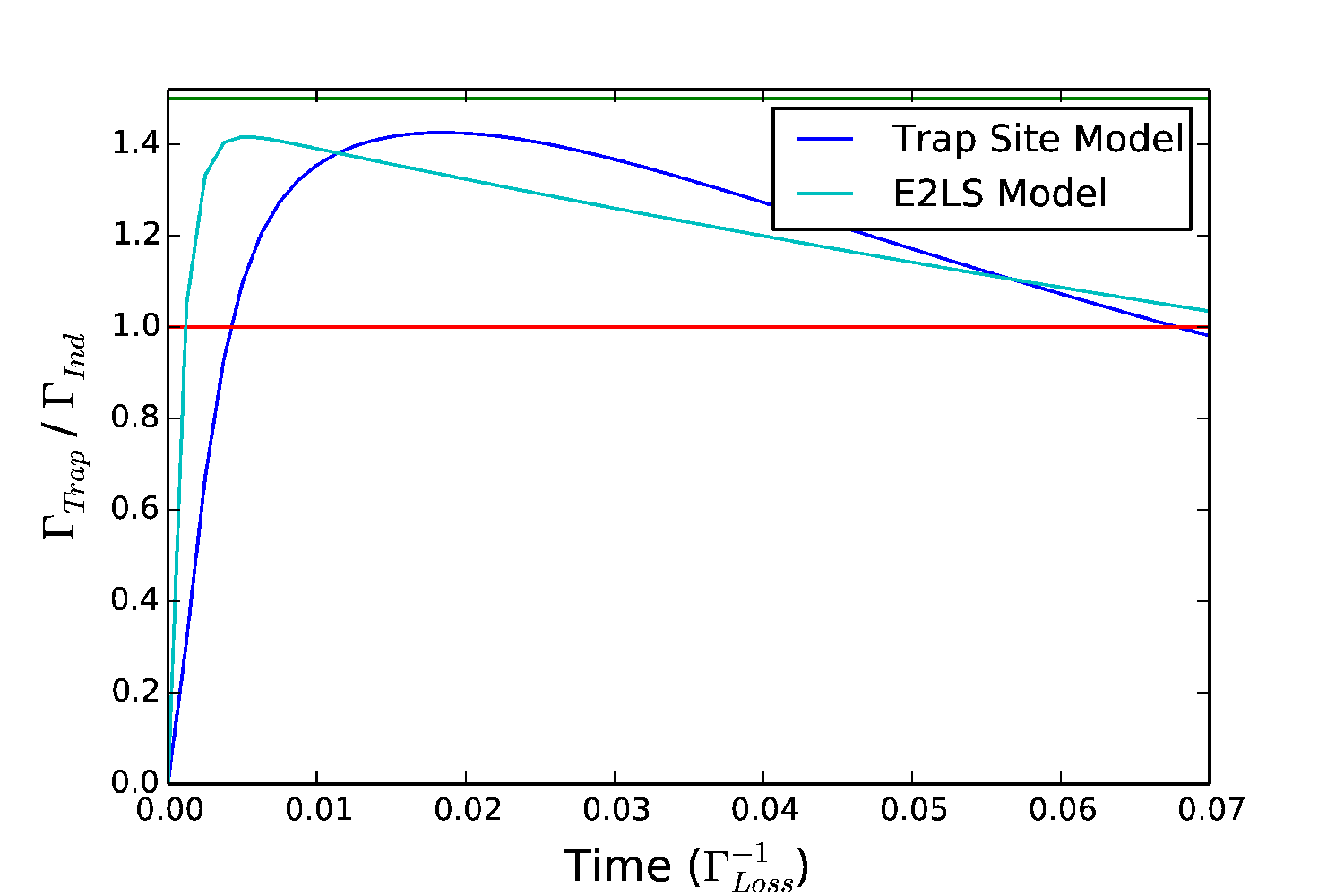}
\caption{ \label{fig_2} Comparison between the full trapping model discussed in Section \ref{sec_trap} modelled numerically using the technique described in \ref{sec_nme} and the effective two level system model with adjusted rates. Parameters are: $d=0.1, \gamma = 10^{-4},  g= 1$, and $\Gamma_{\rm trap} = 4g$. In the phenomenological case $\Gamma_{\rm trap}$ takes the same value, whereas the effective $\Gamma_{\rm loss}$ rate was increased by a factor of five to account for additional leakage out of the E2LS due to detuned exciton extraction processes (see Fig.~\ref{fig:5}). Note that the shape of the E2LS curve can be made to look more similar to the full model by adjusting its $\Gamma_{\rm trap}$ and $\Gamma_{\rm loss}$ rates, but here we have chosen values giving rise to a similar peak height and enhancement area, indicative of a comparable collective advantage over one superabsorption lifetime cycle. }
\end{figure}

\begin{align}\label{E2LS}
\rho_{\text{M}=0}(t) = & \frac{\Gamma_{\rm absorb}}{\sqrt{\Gamma_{\rm total}^2-4\Gamma_{\rm loss}(\Gamma_{\rm emit}+\Gamma_{\rm trap})}} \times \\ \nonumber
&(e^{-\frac{1}{2}t(\Gamma_ {\rm total}-\sqrt{\Gamma_{\rm total}^2-4\Gamma_L(\Gamma_{\rm emit}+\Gamma_{\rm trap})})} -e^{-\frac{1}{2}t(\Gamma_ {\rm total}+\sqrt{\Gamma_{\rm total}^2-4\Gamma_L(\Gamma_{\rm emit}+\Gamma_{\rm trap})})}), \\
 I_{\text{trap}}(t)= & \Gamma_{\rm trap} \, \rho_{\text{M}=0}(t),
\end{align}
where $\Gamma_ {\rm total}  =  \Gamma_{\rm absorb}+\Gamma_{\rm emit}+\Gamma_{\rm loss}+\Gamma_{\rm trap}$ and with $\Gamma_{\rm trap}$ referring to the effective rate of a trapping Lindblad operator, which is related, but generally not necessarily equal to decay rate of the trap site in the full model. This is due to the effect additional parameters such as the coupling strength $g$. By accounting for this and the additional contribution to $\Gamma_{\rm loss}$ described above, we can compare the trap site model against the E2LS.  Figure \ref{fig_2} shows the results of this comparison. The analytical results for the E2LS can thus provide an adequate and simple qualitative description of the dynamics that is brought about by a more complex and realistic trapping model.

\section{Numerical Models and the Effect of  Disorder} \label{sec:num_res}

In this Section we shall verify the result of the main text using numerical approaches. As a first step, we investigate the validity of the master equation framework described in the main paper. This is achieved with a Monte Carlo approach which allows the system to explore the entire Dicke ladder. In order to study disorder in the system Hamiltonian, we also present results from a completely independent model, which only builds on the physical Hamiltonian and makes no assumptions about the Dicke model being a good description of the system.

\begin{figure}[t]
\includegraphics[scale=0.5]{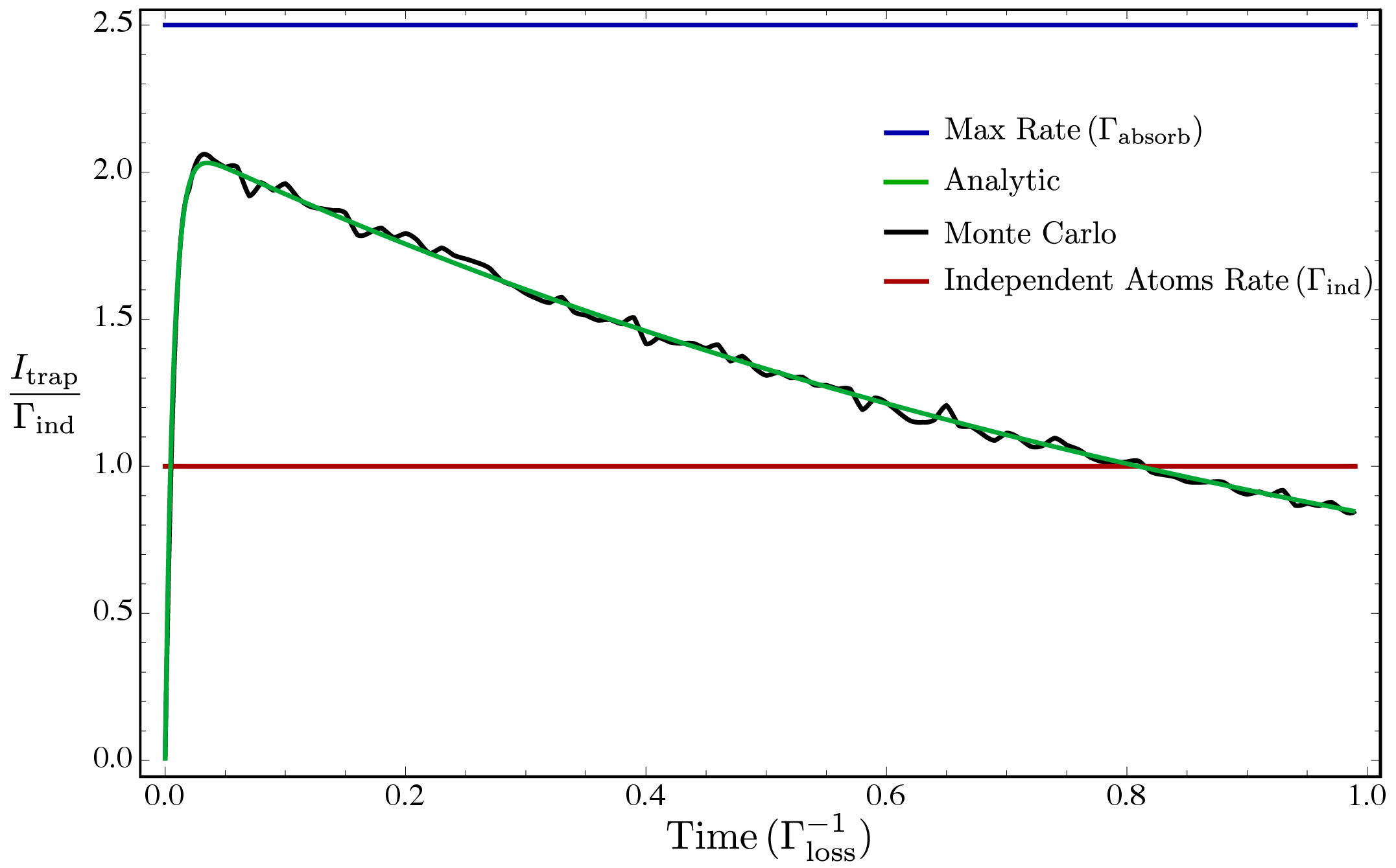}
\caption{ \label{fig_5} Comparison of the Effective Two Level System (E2LS) model with numerical results from Monte Carlo simulations done using the Quantum Optics Toolbox in Python (QuTip). Parameters: $N=8$, $N(\omega_{\rm{good}})= 10$, $\Gamma_{T}= 10\, \Gamma_{E}$ ,$\gamma=1$, trajectories 100,000. }
\end{figure}

\subsection{Monte Carlo}

\begin{figure}[t]
\includegraphics[scale=0.62]{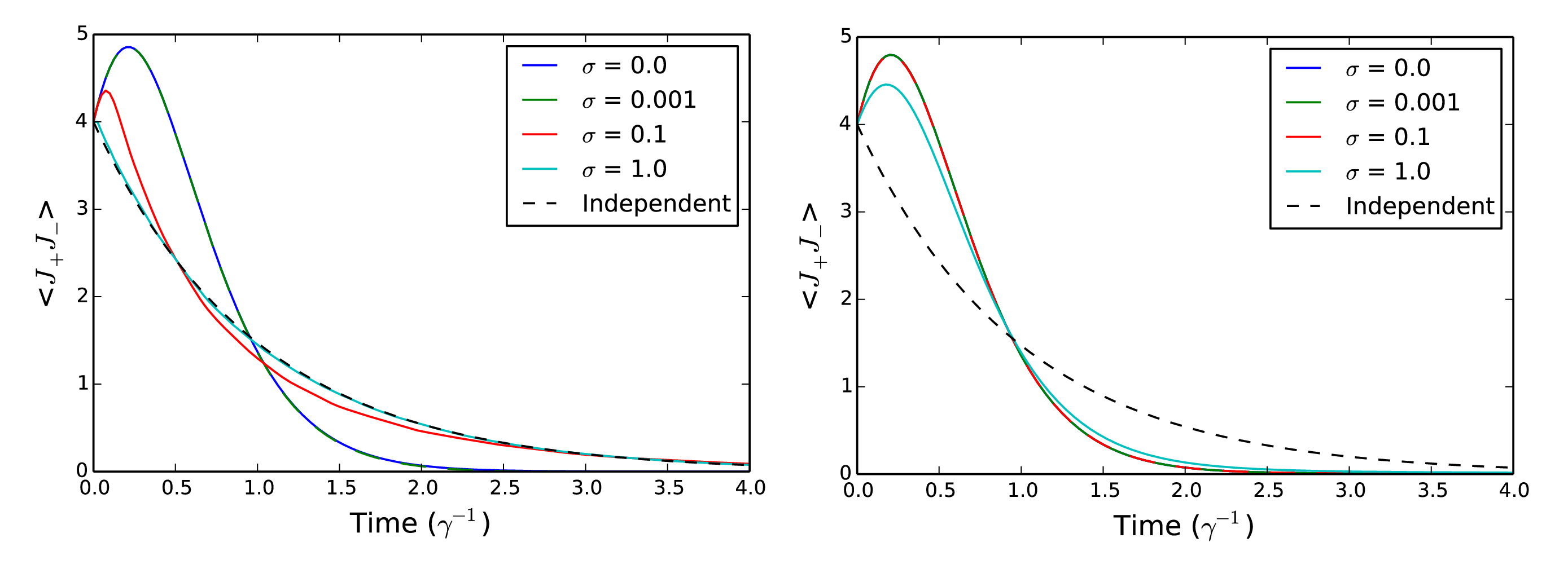}
\caption{ \label{fig_3} Left: The effect of  increasing disorder (modelled as a Gaussian distribution with standard deviation $\sigma$) on a superradiant system without interactions ($\Omega=0$). Right: The effect of  increasing disorder ($\sigma$) on a superradiant system with hopping interaction strength $ \Omega=-1$. Shared parameters: $\omega_{A}=10, d=1, \gamma = 0.01, N=4$. Without interactions, the relevant energy scale characterising the transition from collective to independent emission is given by $\sigma / \gamma$. By contrast, when interactions are included the system eigenstates are intrinsically delocalised and the relevant energy scale becomes $\sigma / \Omega$, leading to a significant increase in robustness against disorder.}
\end{figure}

The Effective Two Level System (E2LS) model in the main paper allowed us to reduce the complexity of the problem dramatically and made it analytically tractable. In order to verify this approach we here compare it to an independent numerical model: Figure \ref{fig_5}  shows excellent agreement  between the E2LS model and Monte Carlo simulations of the master equation (\ref{MEgen_Apen}) [Eq.~(7) of the Article] using QuTip \cite{johansson2013}. Monte Carlo simulations of quantum systems use a time/memory tradeoff to allow large systems to be simulated. They do this by only storing and propagating the state vector rather than the entire density matrix, by averaging over many trajectories the results converge toward the dynamics that would have been obtained by numerical integrating the master equation as in the previous section. The numerical model uses a phenomenological trapping. The agreement with the E2LS can be made arbitrarily close with increasing numbers of trajectories. The same agreement was also seen for systems with larger numbers of atoms. 

\subsection{Numerical integration of Master Equation}\label{sec_nme}

We proceed by exploring the effect of disorder on both superradiance and superabsorption. To capture these effects we must use a more general, but less analytically informative approach to deriving the master equation than the one presented in the main paper and derived in detail in Section \ref{ME_deriv} of this document.  

Starting with the implicit interaction model, we now allow the site energies to vary. We aim  to simulate the effect of static energy disorder, which  -- to some extent -- will be present in all physical implementations of  superabsorption. To do this we draw the site frequencies $\omega_{m}$ from a Gaussian distribution with a mean given by the atomic frequency $\omega_{A}$ and a standard deviation given by $\sigma$. The system Hamiltonian is:

\begin{equation}\label{HamiltonianS_dis}
\hat{H}_{S} =  \sum_{m=1}^{N} \omega_{m} \hat{\sigma}^{m}_{+}\hat{\sigma}^{m}_{-}+  \Omega \sum_{i, j}^{N}(\hat{\sigma}_{+}^{i} \hat{\sigma}_{-}^{j}+ \hat{\sigma}_{-}^{j} \hat{\sigma}_{+}^{i}) \, .
\end{equation}
If the full trap model is to be used the extra terms (\ref{H_trap}) are added. The Hamiltonian is then diagonalised numerically and the eigenoperators determined which project the dipole operator on the eigenspace of $\hat{H}_{S}$: 

\begin{equation} \label{projector}
\hat{A}(\omega) = \sum_{e,e'} \delta(\omega=e-e') \kb{e}{e}  (\hat{J}_{-}+ \hat{J}_{+}) \kb{e'}{e'},
\end{equation}
where  $\kb{e}{e}$ is a projector onto the a given eigenstate of the Hamiltonian $\label{HamiltonianS}$. $\hat{A}_{\alpha}(\omega)$ is therefore an eigenoperator of $\hat{H}_{S}$, which causes a transition between two eigenstates of $\hat{H}_{S}$ sampling the environment  at frequency $\omega$. We construct the new master equation analogously to Eq.~(\ref{master4}): 

\begin{equation}
\frac{d}{dt}\tilde{\rho}_{\cal{S}}(t)=-i \int_0^\infty{dt'\text{Tr}_{\cal{E}}[\tilde{H}_{I}(t), [\tilde{H}_{I}(t-t'), \tilde{\rho}_{\cal{S}}(t) \otimes \tilde{\rho}_{\cal{E}}]} \, ,
\end{equation}
and the standard procedure \cite{breuer02} straightforwardly yields
\begin{equation} \label{nonRWAME}
\frac{d}{dt} \rho_{S}(t) =  \sum_{\omega, \omega'} \sum_{\alpha, \beta} e^{i(\omega-\omega')t} \Gamma_{\alpha, \beta}(\omega)(\hat{A}_{\beta}(\omega)\rho_{S}(t) \hat{A}^{\dagger}_{\alpha}(\omega')- \hat{A}^{\dagger}_{\alpha}(\omega') \hat{A}_{\beta}(\omega)\rho_{S}(t)) + \mathrm{h.c.} ~.
\end{equation}
Equation (\ref{nonRWAME}) could directly be solved numerically,  but this is extremely computationally expensive even for small disordered systems, due the pathological scaling in the number of terms arising from the summation over $\omega$, $\omega'$, $\alpha$ and $\beta$.  To overcome this difficulty, (\ref{nonRWAME}) is traditionally simplified using the rotating wave approximation (RWA). This means neglecting the non-secular terms (where $\omega \ne \omega'$),  leading to the canonical 'quantum optical master equation':

 \begin{equation} \label{quopME}
\frac{d}{dt} \rho_{S}(t) =  \sum_{\omega} \sum_{\alpha, \beta} \Gamma_{\alpha, \beta}(\omega)(\hat{A}_{\beta}(\omega)\rho_{S}(t) \hat{A}^{\dagger}_{\alpha}(\omega)- \hat{A}^{\dagger}_{\alpha}(\omega) \hat{A}_{\beta}(\omega)\rho_{S}(t)) + \mathrm{h.c.} ~.
\end{equation}

In the case of ordinary superradiance, with no dipole-dipole interaction and consequent shifts,  this step is unproblematic because there is in fact only one transition frequency ($\omega_{A}$). However, for all other cases, care must be taken to apply the approximation selectively only where it is strictly justified. The RWA assumes that the fast oscillating terms $e^{i(\omega-\omega')t}$ effectively average to out to zero over the timescale relevant to the system dynamics $\tau_{R}$. The fastest system dynamics timescale is given by the reciprocal of the lowest non zero transition frequency $\omega^{-1}_{\text{min}}$. However, there is some subtlety to applying the RWA to collective transitions in disordered systems: Neglecting all non-secular terms instantly imposes the independent exponential decay type behaviour on the system. Applying no restriction is extremely computationally expensive,  given that disorder calculation must be repeated many times and averaged to produce meaningful results. Removing too many non-secular terms causes an overestimation of the debilitating effect of disorder on the system:  the dynamics to abruptly change from collective to independent behaviour with even the smallest amount of disorder. This is true of the commonly used oder of magnitude separation between $\tau_{R}$ and $\omega^{-1}_{\text{min}}$. Instead the criterion should made steadily more stringent until convergence is attained. By retaining these terms our model can smoothly interpolate between the limits of collective behaviour (superradiance/superabsorption) and independent exponential decay, which should emerge for strongly disordered systems.  The resulting equation is then integrated numerically using the QuTip project \cite{johansson2013}. The numerical calculation is then repeated many times in order to obtain the an ensemble average over different instances of disorder and smooth out the random oscillations introduced by any given instance. 

\begin{figure}[t]
\includegraphics[scale=0.70]{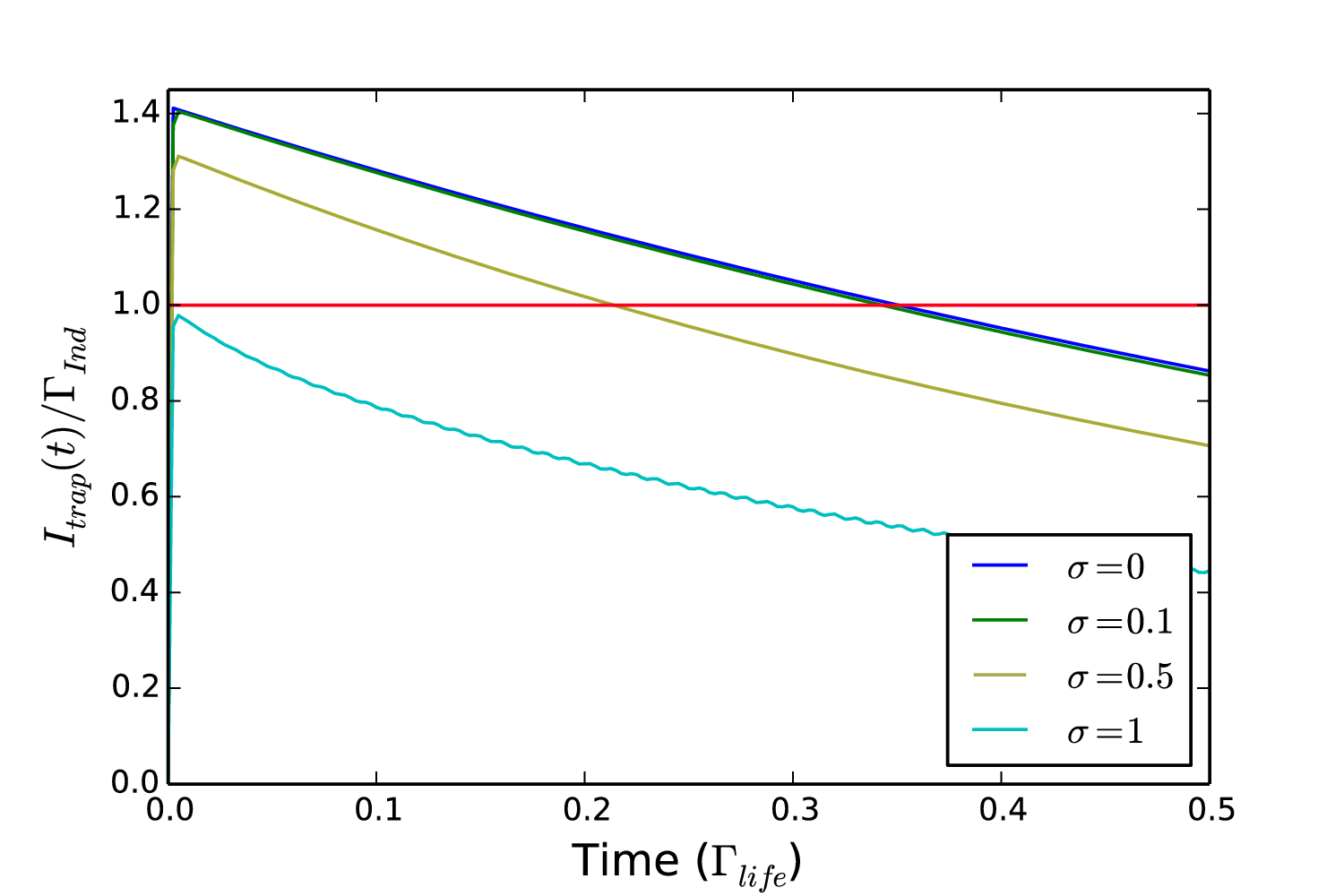}
\caption{ \label{fig_4} Showing the effect of disorder on a superabsorbing system system. Parameters  $\omega_{A}=10$, d=1, $\gamma = 0.01$, N=4., $N(\omega_{\rm{good}})= 1$, $\kappa(\omega_{\rm{bad}})= 0.1\gamma$}
\end{figure}

Figure \ref{fig_3} shows the effect of disorder on superradiance. As expected both plots show a certain robustness to disorder. This is because the increased transition rates produced by superradiance serve to counterbalance the effect of disorder: the faster rate broadens the natural linewidth of the transitions, effectively masking the distinguishability atoms introduced by the disorder. The model with dipole-dipole interactions shows greater robustness, as might be expected because its eigenstates are intrinsically delocalised, whereas with no interactions there is less of a barrier to the localisation introduced by $\sigma$. Figure \ref{fig_4} shows the effect of disorder on superabsorption. As expected from the argument above superabsorption also shows excellent robustness to disorder. It is slightly less robust than superradiance, because it relies on the Dicke ladder shifts not being too heavily altered by the level of static disorder characterised by $\sigma$. The ratio of static disorder to interaction strength needed for superabsorption, is well met by current experimental systems as detailed in the next section.

\section{Material parameters for real systems}

\begin{table}[t!]
\centering 
\begin{tabular}{c c c c c c c} 
\hline \hline 
Type \,\,\,   & E (eV) \,\, & $\gamma^{-1}$ (ns)\,\ & d (D) \,\, & $r_{ij}$  (nm)  \,\, & $\Omega$ (eV)  \,\, & $\sigma$ (meV) \\ [0.5ex] 
\hline 
Quantum Dot (F\"orster)& 1-2  & $ 0.1-1$ & 20-100  & 10-50 &  0.001 &$ \lesssim10$ \\ 
Molecular Ring (B850) &1-2 & $1$ & 5 & $\lesssim 1$ &  0.05 &   $\lesssim 20$\\ 
J-Aggregate &2-3 & $0.05-1$ &10-15 & $\lesssim 1$ &  0.1-0.2 & $5-50$ \\  [0.5ex] 
\hline    
\end{tabular}
\label{tab_mats}  
\caption{Materials parameters comparison: Order of magnitude estimates of the relevant parameters for the superabsorption effect for different systems. Values are taken from the literature, for references see SI text. } 
\end{table}

In this section we consider the parameters for various physical systems that could potentially demonstrate the superabsorption effect. Quantum dots are a good candidate for their large transition dipole moments, continually improving spectral uniformity (low $\sigma$) and the recent progress in synthesising highly ordered rings and arrays \cite{creasey2012}. The main challenge facing this implementation is the relatively weak dipole-dipole interactions ($\Omega$) that have thus far been observed ($\approx 0.01$ meV), although an order of magnitude improvement in this should be easily obtainable \cite{unold2005}. Furthermore, the interaction need not be of the field induced character  this paper focuses on, chosen largely for the sake of simplicity. Any physical mechanism leading to an exciton number dependant shift of the Dicke states would suffice. For instance, in quantum dots the coulomb interaction is known to be stronger ($\sim 10$ meV) \cite{lovett2003}. In general, engineering the strength and the scalability of dot to dot interactions is key a focus for the field, particularly for implementations of quantum information processing tasks. Progress in this area therefore seems very likely in the near future.

Molecular rings have the obvious advantage of possessing the required symmetry and the very close separation between sites, leading to large $\Omega$. The values in the table correspond to the natural photosynthetic ring structure B850. It should be noted that lower disorder would be expected from artificially synthesised rings lacking the protein manifold found in natural systems . Furthermore, the dipole alignment of B850 (in plane) is pessimal for interactions with the field (B850 plays a storage and transfer roll in photosynthesis) and therefore could be expected to increase by a factor of 2 if the dipoles were optimally oriented.  Artificial porphyrin rings could alleviate both of these problems \cite{osullivan2011}. 

J-Aggregates constitute a particularly promising candidate, having both highly  delocalised excitations and very strong interactions between monomers \cite{kobayashi2012}. Their interactions are sufficiently strong to overcome the typical values for disorder \cite{bassani2003}. Collective effects such as superradiance and line narrowing have long been demonstrated in these systems.  It is also possible to  control geometry of J-Aggregates chemically and deposit them on surfaces \cite{saikin2013}. Integrating theses structures with quantum control such as photonic bandgap crystals and optical microcavities is a new and developing area at the confluence of several fields. The objective of this effort is primarily the the study of exciton transport in photosynthesis, but at the same time, these systems provide an extremely promising platform for demonstrating superabsorption.